\documentclass[useAMS,usenatbib]{mn2e}

\usepackage{times}
\usepackage{graphicx}
\usepackage{amsmath}
\usepackage{amssymb}
\usepackage{natbib}
\usepackage{color}
\usepackage{epsfig}
\usepackage{psfig}
\usepackage{journal_names}
\usepackage{paralist}

\newcommand{\kms}{\mbox{\,km~s$^{-1}$}}

\newcommand{\comment}[1]{}

\title[The globular cluster system of NGC~4649]
{A SLUGGS and Gemini/GMOS combined study of the elliptical galaxy M60: wide-field photometry and kinematics of the globular cluster system}
\author[Pota~et~al.~ ]
{Vincenzo Pota$^{1}$, Jean P. Brodie$^{1}$, Terry Bridges$^{2}$, Jay Strader$^{3}$, Aaron J. Romanowsky$^{1,4}$,\\
\\
\normalfont{\LARGE  Alexa Villaume$^1$, Zach Jennings$^1$, Favio R. Faifer$^{5,6}$, Nicola Pastorello$^7$, Duncan A. Forbes$^{7}$}  \\
\\
\normalfont{\LARGE Ainsley Campbell$^2$, Christopher Usher$^7$, Caroline Foster$^8$, Lee R. Spitler$^{8,9}$, Nelson Caldwell$^{10}$} 
\\
\\  
\normalfont{\LARGE Juan C. Forte$^5$, Mark A. Norris$^{11}$, Stephen E. Zepf$^3$,  
Michael A. Beasley$^{12,13}$, Karl Gebhardt$^{14}$} 
\\
\\
\normalfont{\LARGE David A. Hanes$^{2}$, Ray M. Sharples$^{15}$, Jacob A. Arnold$^1$} 
\\    
\\
$^{1}$ University of California Observatories, 1156 High Street, Santa Cruz, CA 95064, USA\\
$^{2}$ Department of Physics, Engineering Physics, and Astronomy, Queen's University, Kingston, ON K7L 3N6, Canada\\
$^{3}$ Department of Physics and Astronomy, Michigan State University, East Lansing, Michigan 48824, USA\\
$^{4}$ Department of Physics and Astronomy, San Jos\'e State University, One Washington Square, San Jose, CA 95192, USA \\
$^{5}$ Consejo Nacional de Investigaciones Cientficas y T\'ecnicas (CONICET)-Planetario Galileo Galilei, CABA, Rep. Argentina\\
$^{6}$ Instituto de Astrof\'isica de La Plata (CCT La Plata - CONICET - UNLP) \\
$^{7}$ Centre for Astrophysics \& Supercomputing, Swinburne University, Hawthorn VIC 3122, Australia\\
$^{8}$ Australian Astronomical Observatory, PO Box 915, North Ryde, NSW 1670, Australia\\
$^{9}$ Department of Physics \& Astronomy, Macquarie University, Sydney, NSW 2109, Australia\\
$^{10}$ Harvard-Smithsonian Center for Astrophysics, Cambridge, MA, USA\\
$^{11}$ Max Planck Institut f\"{u}r Astronomie, K\"{o}nigstuhl 17, D-69117 Heidelberg, Germany\\
$^{12}$ University of La Laguna. Avda. Astrof\'isico Fco. S\'anchez, s/n. E38206, La Laguna, Tenerife, Canary Islands, Spain.\\
$^{13}$ Instituto de Astrof\'\i sica de Canarias. Calle V\'\i a L\'actea s/n. E38200 - La Laguna, Tenerife,
Canary Islands, Spain.\\
$^{14}$ Astronomy Department, University of Texas, Austin, TX 78712, USA\\
$^{15}$ Department of Physics, University of Durham, South Road, Durham DH1 3LE }

\date{March 2015}

\begin{document}

\label{firstpage}

\maketitle

\begin{abstract}
We present new wide-field photometry and spectroscopy of the globular clusters (GCs) around NGC~4649 (M60), the third brightest galaxy in the Virgo cluster. Imaging of NGC~4649 was assembled from a recently-obtained \textit{HST}/ACS mosaic, and new Subaru/Suprime-Cam and archival CFHT/MegaCam data. About 1200 sources were followed up spectroscopically using combined observations from three multi-object spectrographs: Keck/DEIMOS,
Gemini/GMOS and MMT/Hectospec. We confirm 431 unique GCs belonging to NGC~4649, a factor of 3.5 larger than previous datasets and with a factor of 3 improvement in velocity precision. 
We confirm significant GC colour bimodality and find that the red GCs are more centrally concentrated, while the blue GCs are more spatially extended. We infer negative GC colour gradients in the innermost 20 kpc and flat gradients out to large radii. Rotation is detected along the galaxy major axis for all tracers: blue GCs, red GCs, galaxy stars and planetary nebulae.
We compare the observed properties of NGC~4649 with galaxy formation models. We find that formation via a major merger between two gas-poor galaxies, followed by satellite accretion, can consistently reproduce the observations of NGC~4649 at different radii. We find no strong evidence to support an interaction between NGC~4649 and the neighbouring spiral galaxy NGC~4647. We identify interesting GC kinematic features in our data, such as counter-rotating subgroups and bumpy kinematic profiles, which encode more clues about the formation history of NGC~4649.
\end{abstract}

\section{Introduction}

Galaxies are the building blocks of the visible Universe and the best tools to study its structure. Galaxies are alike in many ways, suggesting that similar underlying mechanisms regulate their formation, but they are not identical to each other, implying that formation processes vary slightly from galaxy to galaxy \citep{Lintott08}. While high redshift galaxy surveys can be used to infer ``average'' formation mechanisms for a sample of galaxies, studying nearby galaxies in detail can allow us to reconstruct their particular and unique formation histories. Moreover, observations of galaxies at $z=0$ provide the end product that computer simulations attempt to reproduce \citep[e.g.,][]{Hoffman,Wu12,Naab14}.

In the standard galaxy formation theory, small structures collapse first \citep{White78,Searle78,Zolotov}, and then grow hierarchically into larger structures via galaxy mergers and/or via accretion of satellite galaxies \citep[e.g.,][]{Lopez10,Font,vanDokkum14,Tasca14}. The latter is responsible for the building-up of galaxy outskirts (which we will refer to as stellar haloes) from high ($z\approx2$) to low redshift \citep{Tal11,Oser,Khochfar11}. The various processes which shaped galaxies with time can be simulated and compared with observations of galaxies at a particular redshift. For example, minor mergers can create stellar shells and tidal streams observable in deep imaging \citep{Tal,Ebrova13,Atkinson13,Duc15}, whereas the secular accretion of satellite galaxies can be studied with photometry of stacked galaxies \citep[e.g.,][]{Oser,Tal11,DSouza14} or by studying the chemistry and kinematics of stellar haloes \citep[e.g.,][]{Forbes11,Romanowsky12,Coccato13}.

Galaxy haloes are difficult to study because they are optically faint. Globular clusters (GCs), on the other hand, are much more observationally convenient for studying galaxy haloes. They are relatively easy to detect because of their high surface brightness and they generally populate haloes in large numbers \citep[e.g.,][]{Brodie}. The outermost GCs can be used to map out halo properties (e.g., kinematics and metallicity) out to tens of effective radii \citep[e.g.,][]{Ostrov,Schuberth,Usher12,Pota13,Forbes11}. Their old ages and their direct connection with the star forming episodes in a galaxy's history \citep{Strader05,Puzia05} mean that they are ideal tracers of  the assembly processes that shaped the host galaxy \citep[e.g.,][]{Romanowsky12,Leaman13,Foster14,Veljanoski14}. 

In this paper we study the GC system of the E2 early-type galaxy NGC~4649 (M60), the third brightest galaxy in the Virgo cluster. NGC~4649 is itself at the centre of a small group of galaxies and is the dominant member of the galaxy-pair Arp~116 \citep{Arp66}. The proximity to the disturbed spiral galaxy NGC~4647 (13 kpc in projection) potentially makes NGC~4649 an example of a pre major-merger between two massive galaxies in the local Universe, although this connection is still debated \citep{Young06,deGrijs}.

Galaxies like NGC~4649 are ideal laboratories to test galaxy formation models. This galaxy has been scrutinized from different angles, revealing the portrait of a prototypical red and dead elliptical galaxy with a central supermassive black hole \citep{Shen10}, dark matter halo \citep{Bridges06,Das11} and X-ray halo \citep{OSullivan,Das10,Humphrey08,Paggi14}.

Recent space-based observations of NGC~4649 \citep{Strader12,Luo13,Norris14}, suggest that this galaxy interacted with a smaller galaxy with mass $M\sim10^{10} M_{\sun}$ \citep{Seth14}.
This event should have deposited $~100$ GCs and many more stars into the halo of NGC~4649 during the course of its orbit \citep{Harris13}. Indeed, asymmetries have been found in the two-dimensional distributions of GCs and low-mass X-ray binaries around this galaxy \citep{Mineo14,DAbrusco14}, but these have not yet been linked to any particular interaction that occurred in NGC~4649. 

In addition to recent space-based observations, the GC system of NGC~4649 has been studied with ground-based telescopes \citep{Bridges06,Pierce4649,Lee08,Faifer11,Chies-Santos}. A spectroscopic follow up of 121 GCs \citep{Hwang08} revealed that the GC system of NGC~4649 has a large rotation amplitude. This is a rare feature in large early-type galaxies, whose GC systems are generally pressure supported with negligible or weak rotation \citep[e.g.,][]{Cote03,Bergond06,Schuberth10,Strader11,Norris12,Pota13,Richtler14}, although exceptions exist \citep[e.g.,][]{Puzia04,Arnold,Schuberth12,Blom}. On the other hand,
\citet{Bridges06} obtained radial velocities for 38 GCs in NGC~4649 and found no rotation, probably because of their relatively small sample size. 

In this paper we present the largest spectro-photometric catalogue of GCs around NGC~4649. 
We construct a wide-field photometric GC catalogue by combining literature (\textit{HST}), archival (CFHT) and new (Subaru) observations. We follow-up with spectroscopy of hundreds of GC candidates using joint observations from three multi-object spectrographs mounted on the Keck, MMT and Gemini telescopes. Our new spectroscopic catalogue is a factor of 3.5 larger than the current literature dataset \citep{Lee4649} and has a factor of 3 greater velocity accuracy. 
This study exploits datasets from two complementary surveys of extragalactic GCs: the SLUGGS survey \citep{Brodie14}, and an ongoing survey carried out with Gemini/GMOS (Bridges et al., in preparation). These two surveys combined, contribute 90 per cent of the final sample of confirmed GCs.
We use the photometric and kinematic properties of the NGC~4649 GC system (including the blue and red GC subpopulations) to study the formation history of NGC~4649. A follow-up paper (Gebhardt et al., in preparation) will make use of the dataset presented in this work to model the mass content of NGC~4649. 

We adopt a distance of 16.5 Mpc \citep{Mei07}, an effective radius $R_e= 66$ arcsec $ = 5.3$ kpc, an axis ratio $q=0.84$ and a position angle $PA=93$ deg \citep{Brodie14}. Galactocentric distances $R$ are expressed though the circularized radius $R=\sqrt{X^2 q  + Y^2 /q}$, where $X$ and $Y$ are the Cartesian coordinates of an object with NGC~4649 at the origin. The absolute magnitude of NGC~4649 is $M_B=-21.47$ mag or $M_V=-22.38$ mag.

This paper is structured as follows.
Imaging observations and analysis are presented in Sections \ref{sec:imaging} and \ref{sec:imaginganalysis} respectively. The photometric results, including colour gradients and GC surface density are presented in Section \ref{sec:photresults}. The spectroscopic observations and their outcomes are discussed in Section \ref{sec:spectroscopy} and Section \ref{sec:specresults}, respectively. We next explain the steps needed to build up the spectroscopic GC master catalogue: repeated GC measurements (Section \ref{sec:repeated}) and colour and velocity uncertainty calibration (Section \ref{sec:master}). We give an overview of the spectroscopic sample in Section \ref{sec:overview} and explain the tools to quantify GC kinematics in Section \ref{sec:kinematics}. The kinematic modelling results are given in Section \ref{sec:kineresults}. We discuss our results in Section \ref{sec:discussion} and summarize the paper in Section \ref{summary}.

\begin{figure*}
\centering
\includegraphics[trim = 60mm 25mm 10mm 1mm,clip, scale=0.7]{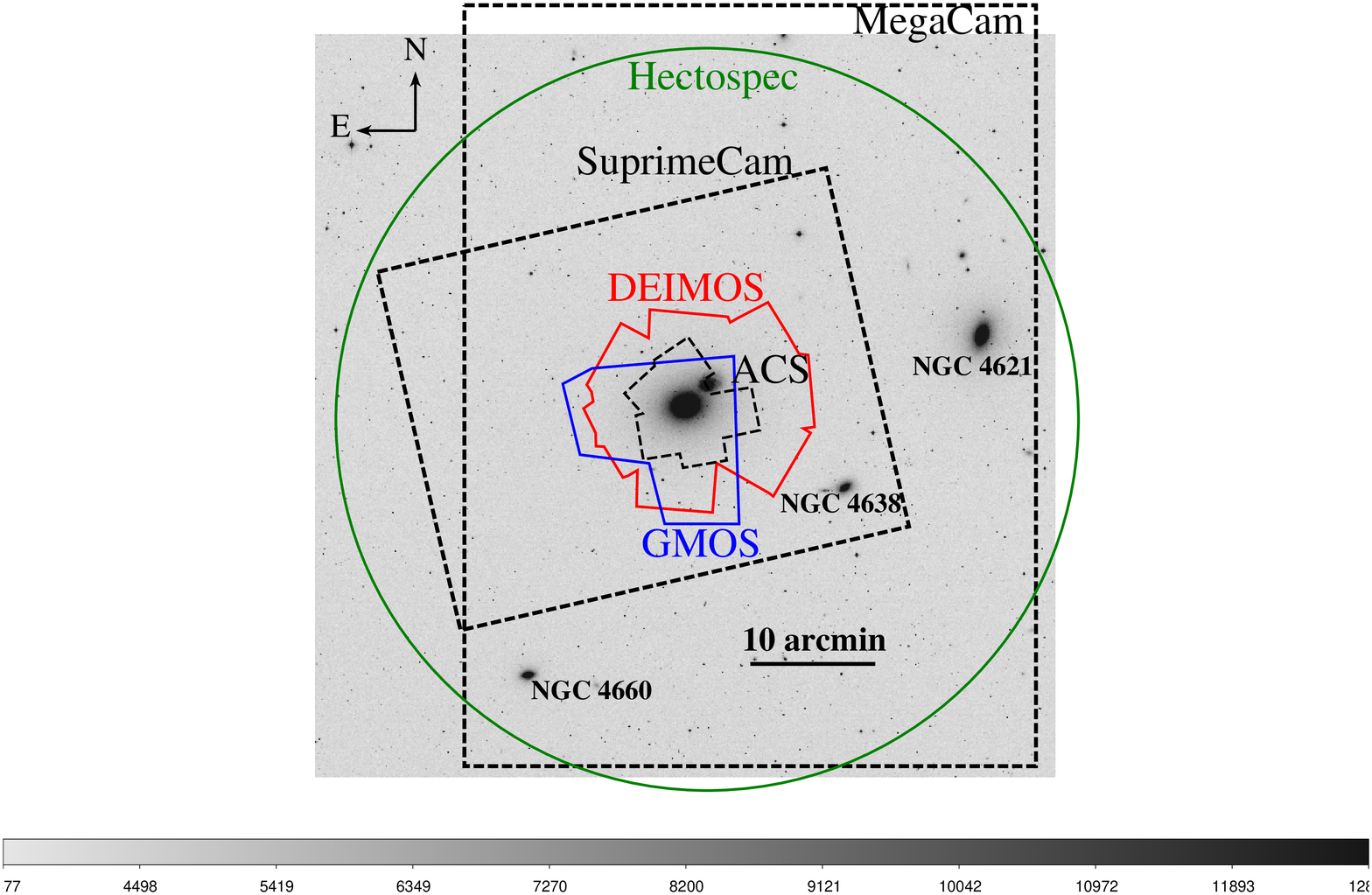} 
\caption{Overview of the observations. Shown is a Digitized Sky Survey image of NGC~4649 and surroundings. The field of view is 1 deg$^2$, which corresponds to 290 kpc $\times$ 290 kpc at this distance. NGC~4649 is the elliptical galaxy at the centre of the image, with the spiral NGC~4647 visible on the immediate right. Other galaxies in the field are labelled. Black dashed polygons represent the fields-of-view of the imaging cameras (labelled accordingly). Coloured solid outlines enclose the regions in the sky mapped by our three multi-object spectrographs (labelled). A 10 arcmin$ \sim 48$ kpc scale-bar is shown at the bottom of the image. North is up, East is left.}
\label{fig:imag}
\end{figure*}

\section{Imaging observations}
\label{sec:imaging}

Imaging is the first step in identifying extragalactic GCs. It allows us to identify and characterize the photometric properties of the NGC~4649 GC system, and also to prune out the bulk of the contaminants (Galactic stars and background galaxies). Imaging is also necessary for selecting objects to be followed up spectroscopically.

Our imaging comes from three sources: \textit{HST}/ACS, CFHT/MegaCam and Subaru/Suprime-Cam. The fields of view of the three instruments, along with the footprints of the multi-object spectrographs, are shown in Figure~\ref{fig:imag}.
CFHT/MegaCam covers the field of interest, and it was supplemented with Subaru/Suprime-Cam data to gain extra coverage on the East of NGC~4649 and to test for systematics in our ground-based photometry.  

\subsection{\textit{HST}/ACS}
\label{sec:ACS}

The \textit{HST} imaging is discussed in \citet{Strader12} and we refer the reader to this paper for a detailed description of the data. The dataset consists of six ACS pointings, which extend up to 6 arcmin from the galaxy center. The layout of the pointings is such that it avoids the foreground spiral NGC~4647. GC selection is based on $(g - z)$ colours, $z$ magnitudes and half-light radii $r_h$. The latter can be measured down to $\sim 0.1 \times$ FWHM \citep{Spitler06}, which corresponds to $\sim$ 0.7 pc at the distance of NGC~4649. Unresolved sources were not selected as GCs. The latter is the physical radius which contains half of the total light of the object. The final HST/\textit{ACS} catalogue consists of 1603 GC candidates. 

\subsection{CFHT/MegaCam}
\label{sec:CFHT}

The area surrounding NGC~4649 has been imaged with CFHT/MegaCam as part of the Next Generation Virgo Survey \citep{Ferrarese12}. MegaCam has a field-of-view of $0.96\times0.94$ deg$^2$ and a pixel scale of 0.187 arcsec/pixel. 
Reduced CFHT/MegaCam images and the respective weight-images in $gri$ filters were downloaded from the \textsc{CFHT/MegaPipe} legacy archive \citep{Gwyn08}. Images are provided already calibrated to standard SDSS filters. The logs of the CFHT observations are given in Table \ref{tab:imaging}. The seeing for all images is  $<0.8$ arcsec.

We use a total of five MegaCam pointings. Four of these are in $ugiz$-filters, but they are not centered on NGC~4649. A fifth pointing in the $r$-filter is centered on NGC~4649. The final CFHT field of interest spans roughly 1 degree in declination and half a degree in right ascension, as shown in Figure \ref{fig:imag}. We discard 30 arcmin in right ascension as we are not interested in the sky to the West of NGC~4621.  

\subsection{Subaru/Suprime-Cam}
\label{sec:Subaru}

NGC~4649 was imaged with Subaru/Suprime-Cam in $gri$ filters. The field-of-view of Suprime-Cam is $34\times27$ arcmin$^2$, with a pixel scale of 0.2 arcsec/pixel \citep{Miyazaki02}. The camera was pointed at roughly the centre of NGC~4649. A summary of the Subaru observations is given in Table \ref{tab:imaging}. The seeing was sub-arcsec for the $g$ and $i$ images, and 1.1 arcsec for the $r$ image.

We use a custom made pipeline for reducing the Subaru Suprime-Cam imaging, based on the SFRED2 pipeline of \citet{Ouchi}. The reduction consists of the following steps:

\begin{itemize}
\item We subtract the bias measured from the overscan regions for each chip. We create the flat field frames by median combining the dome-flats for each chip. We use L.A. Cosmic \citep{vanDokkum01} to remove the cosmic rays from the object frames. 

\item We correct for chip-to-chip quantum efficiency variations and scale individual chips for any time-variable gain. We correct for the atmospheric distortion, mask out the auto-guider probe in each frame and sky subtract using a simple median of medians. 

\item We use a Monte Carlo sampling method to efficiently find the astrometric solution which minimizes the positional offsets between SDSS and point sources in our data.

\item We multiplicatively scale each exposure to the median throughput level of all the exposures before co-adding them into a final image. The final mosaic image for each filter was performed with the software Montage~\footnote{http://montage.ipac.caltech.edu/index.html}.
\end{itemize}

\begin{table}
\centering
\label{mathmode}
\begin{tabular}{@{}l c c c c}
\hline
Obs ID-Filter & Date & Exposure time & Seeing  \\
\hline
& [HST] & [seconds] & [arcsec]\\
\hline
Subaru--$g$ & 2010-04-11 & 1160 & 0.8 \\ 
Subaru--$r$ & 2010-04-11 & 1260 & 1.1 \\ 
Subaru--$i$ & 2010-04-11 & 490 & 0.7 \\  
\hline
G008.190.711+11.596--$r$ & 2008-03-01 & 2160 & 0.8 \\ 
NGVS+3+0--$g$ & 2009-02-27 & 3170 & 0.6\\ 
NGVS+3+0--$i$ & 2009-02-27 & 2055 & 0.6\\ 
NGVS+3-1--$g$ & 2009-02-27 & 3170 & 0.6\\ 
NGVS+3-1--$i$ & 2009-02-27 & 2055 & 0.6\\ 
NGVS+2+0--$g$ & 2009-02-25 & 3170 & 0.6\\ 
NGVS+2+0--$i$ & 2009-02-15 & 2055 & 0.6\\ 
NGVS+2-1--$g$ & 2009-05-25 & 3170 & 0.6\\ 
NGVS+2-1--$i$ & 2009-06-20 & 2055 & 0.6\\ 
\hline
\end{tabular}
\caption{Ground-based photometric observations. Listed are the telescope and filter, the observation date in Hawaiian Standard Time (HST), the exposure time and the seeing. The overall seeing conditions are excellent.}
\label{tab:imaging} 
\end{table}

\section{Imaging analysis}
\label{sec:imaginganalysis}
\subsection{Catalogue extraction and calibration}

We use SExtractor \citep{Bertin} to extract sources from the Suprime-Cam and MegaCam images. We extract sources from the two cameras separately, and we combine them later on. We run SExtractor on all the fields listed in Table \ref{tab:imaging}. Only sources above $2.5\sigma$ are extracted. The zero point magnitude is set to 30 mag for the CFHT images \citep{Ferrarese12}, and to 25 mag for the Subaru images. The weight-maps downloaded from the MegaPipe archive were used to optimise the extraction of MegaCam objects. The weight-map for the Suprime-Cam dataset was set to ``background'', meaning that we let SExtractor compute the variance map from the science images themselves. 

For each extracted source, we measure: 
\begin{inparaenum}[\itshape a\upshape)]
\item coordinates;
\item \texttt{magbest}, which we use as our primary magnitude measurement;
\item a set of magnitudes computed within different apertures (from 4 to 11 pixels in diameter);
\item structural parameters, such as the ellipticity, position angle and FWHM.
\end{inparaenum}

The photometric properties of the objects in common between the different $g$ and $i$ MegaCam pointings are in good agreement with each other. Selecting objects brighter than $i=23$ mag, we find the median magnitude difference to be $\Delta g = 0.006$ mag and $\Delta i = 0.008$ mag, respectively. The extracted properties for these objects are averaged together so that every object is unique, with magnitude measurements in all three filters.
  
Point sources are selected based on the difference between the magnitude measured within 4 pix and within 8 pix. This difference is small for point sources (typically $<0.6 \mbox{ mag}$) because their light profiles fall off more rapidly than those of extended sources. The latter become important source of contamination for $(i>22 \mbox{ mag})$.

With the point-source catalogues in hand, we calibrate the magnitudes to standard SDSS $gri$ filters. We select a bright sub-sample ($18<i<21$ mag) and match these sources to the SDSS DR7 point-source catalogue, which covers the entire field-of-view in Figure \ref{fig:imag}. 

The MegaCam dataset was already calibrated to SDSS magnitudes and, in fact, we find the zero-point corrections to be always $<0.05$ mag, within the photometric uncertainties. For the Suprime-Cam dataset we find: $g_{\rm SDSS} - g_{\rm SUB} = 3.46 \pm 0.05$ mag, $r_{\rm SDSS} - r_{\rm SUB} = 3.76 \pm 0.05$ mag, $i_{\rm SDSS} - i_{\rm SUB} = 3.66 \pm 0.04$ mag, respectively.

\subsection{Comparison between CFHT and Subaru}

Before merging the MegaCam and Suprime-Cam datasets, we study how the magnitudes extracted from both cameras compare with each other. We match the two catalogues selecting only bright ($19<i<23$) objects outside $R=300$ arcsec, because ground-based imaging is notoriously incomplete near the centres of the galaxies.
 
\begin{figure}
\centering
\includegraphics[width=\columnwidth]{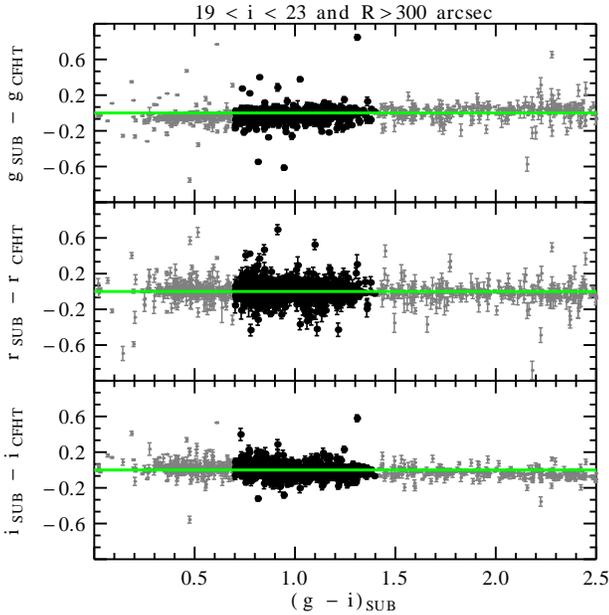} 
\caption{Comparison between Subaru and CFHT datasets. Top, central and bottom panel shows the magnitude difference in $g$, $r$, $i$ filters, respectively, of the point-sources in common between Subaru and CFHT. The green line marks zero magnitude difference. Only objects with $19<i<23$ and $0<(g-i)<2.5$ are shown. Globular clusters fall in the colour range $0.07\le (g-i)\le 1.4$, which is highlighted with black points. In this colour range, the CFHT and Subaru datasets agree with each other. The deviation of $\sim 0.07$ mag for $g$ and $i$ filters, and $\sim 0.09$ mag for the $r$ band.}
\label{fig:SUB_CFHT}
\end{figure}

Figure \ref{fig:SUB_CFHT} shows the magnitude difference in three bands of the $\sim 1800$ point sources in common between MegaCam and Suprime-Cam. We are mostly interested in objects with $0.65\le (g-i)\le 1.4$ because this is the colour range populated by extragalactic GCs, as we will discuss below. 
In this specific colour-magnitude range, the root-mean-square (rms) of the magnitude difference is rms$(\Delta g) =0.070$ mag, rms$(\Delta r) =0.095$ mag, rms$(\Delta i) =0.064$ mag for the $g$, $r$, $i$ filters respectively. The larger value of rms$(\Delta r)$ is due to the relatively poor quality of the Suprime-Cam $r$-band image. 
No significant trend with $(g-i)$ colours is observed in the colour range under investigation. We conclude that the magnitudes measured from MegaCam and from Suprime-Cam are in good agreement. 

We then combine the point-source catalogues from MegaCam and Suprime-Cam. If an object is present in both datasets, we weight-average its magnitudes in the three filters. If an object is present in only one catalogue, it is added to the final master catalogue without applying any magnitude correction. Lastly, we correct the final $gri$ magnitudes for Galactic extinction using $A_g = 0.087 $ mag, $A_r = 0.060 $ mag, $A_i = 0.045 $ mag, respectively \citep{Schlafly11}.These values vary by 10 per cent across our field-of-view of interest. The combined MegaCam--Suprime-Cam point-source catalogue consists of $\sim 20500$ unique objects, including interlopers.

\begin{figure}
\centering
\includegraphics[width=\columnwidth]{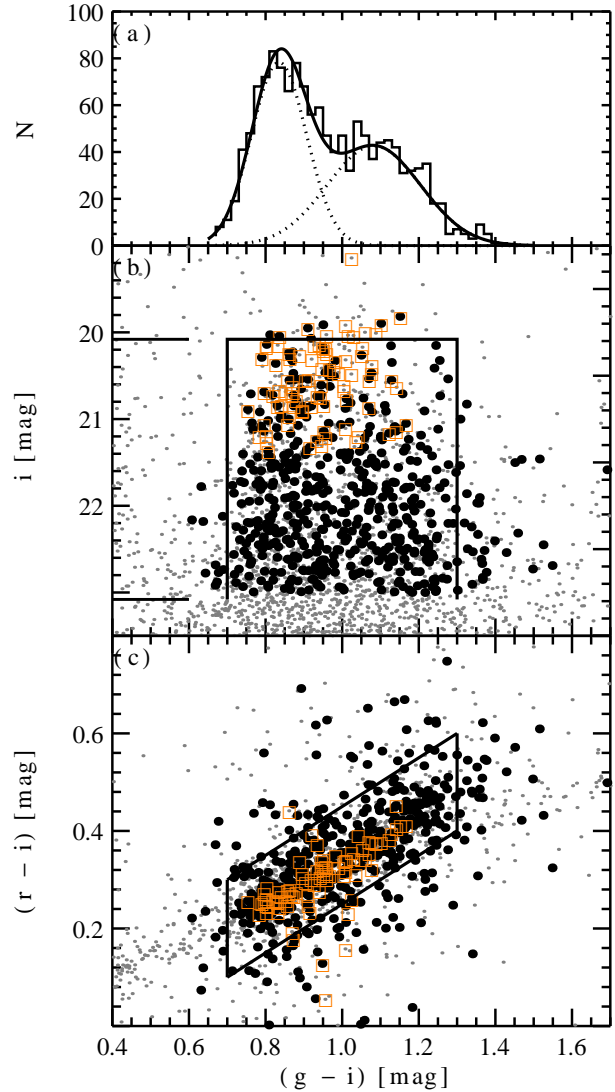} 
\caption{Colour-magnitude and colour-colour magnitude diagrams. For clarity, only ground-based sources within 600 arcsec from NGC~4649 are shown (small grey points). Large black points are partially resolved GCs from HST. Orange open squares are the confirmed GCs from \citet{Hwang08}. The bottom panel shows the colour-colour magnitude plot. The central panel is the colour-magnitude plot. Horizontal lines indicate the TOM of the GC luminosity function $i \mbox{ (TOM)} = 23.1$ mag, and the magnitude of $\omega$ Cen, $i=20$ mag. Objects inside the black lines are flagged as GC candidates. The top panel is the numerical histogram of the selected GCs, with the best double-Gaussian fit to the data.}
\label{fig:CC}
\end{figure}

\subsection{Globular cluster selection}
\label{sec:GCselection}

We use two steps to extract bona-fide GCs from the point-source catalogue. We first exploit the \textit{HST}/ACS catalogue to check where GCs lie in colour-colour and colour-magnitude space, and then we select GCs based on this comparison.

GCs are partially resolved in \textit{HST} images. In NGC~4649 the average GC half-light radii are $r_h\approx$ 2 pc \citep{Strader12}, similar to what is found in other galaxies \citep[e.g.,][]{Masters}. 
This means that even without spectroscopic confirmation, we can flag objects with apparent $r_h < 10$ pc as GCs likely associated with NGC~4649, although some contamination is expected at faint magnitudes. 
From the ACS catalogue, we select GC candidates brighter than $z=23$ mag and with $R>100$ arcsec and we match them with our combined ground-based catalogue.

We find roughly 500 objects in common. Then we look at the colour-magnitude and colour-colour diagrams of the point sources with ground-based photometry, and we highlight these 500 GCs with size confirmation. The result is shown in Figure \ref{fig:CC}. We also matched the ACS catalogue with the 121 spectroscopically confirmed GCs from \citet{Hwang08}, which will be discussed in Section \ref{sec:spectroscopy}. We found 84 objects in common, which are also shown in Figure \ref{fig:CC}. 

NGC~4649 GCs populate a well defined area of both the colour-magnitude and colour-colour diagrams. In the central panel of Figure \ref{fig:CC} we see the well known GC colour bimodality and the blue tilt, which causes the most luminous blue GCs to bend towards redder colors \citep[e.g.,][]{Spitler06,Harris06}. 
At the faint and bright GC selection limits we show, respectively, the expected turn-over magnitude (TOM) of the GC luminosity function ($M_{\rm TOM}, i = - 7.97$ mag; \citealt{Jordan}) and the magnitude of $\omega$ Cen, the brightest star cluster in the Milky Way, for which we assume a magnitude $M_i = -11.0$ mag ($i=20.1$ mag) obtained from \citet{Vanderbeke14} adopting the distance of \citet{vandeVen06}. A handful objects are indeed brighter than this magnitude. We will show that some of these are ultra compact dwarfs (UCDs) with spectroscopic confirmation (see \S \ref{sec:overview}). 

On the bottom panel of Figure \ref{fig:CC} we notice the same grouping of GCs in a narrow colour-colour range. The tail of objects at very red colours are Galactic red giant stars which enter the GC main sequence at $(g-i)\sim1.4$ mag and contaminate the sample. The objects at $(g-i)\le0.6$ are Galactic blue young stars. Roughly 20 per cent of the GC candidates lie outside the selection box. This is due to photometric uncertainties and to the poor quality of ground-based photometry near the centre. Selecting only objects with $z<22$ mag and with $R>200$ arcsec, the fraction of outliers reduces to 3\% .

Objects falling in the colour range highlighted in Figure \ref{fig:CC} are flagged as GC candidates. Objects falling outside the selection box by less than $1\sigma$ are also flagged as GCs. We require a GC to be brighter than the turn-over-magnitude ($i<23.1$ mag) and fainter than $\omega$ Cen. These selection criteria returned $\sim 4000$ GC candidates, with a 14 per cent contamination rate based on the background value which will be computed in Section \ref{sec:photresults}.

The colour distribution of the selected GCs is bimodal (top panel of Figure \ref{fig:CC}). To quantity this bimodality, we use the Kaye's Mixture Model (KMM) algorithm \citep{Ashman94} to fit a double-Gaussian to the GC colour distribution. This calculation was performed only on objects within 600 arcsec from the galaxy centre in order to prune the bulk of contaminants which populate the outer regions of the galaxy (see Section \ref{sec:photresults}).

We find bimodality to be statistically significant with a $p$-value $<10^{-4}$. The blue GC subpopulation is fitted by a Gaussian which peaks at $\mu (g-i)=0.83$ mag, with a dispersion of $\sigma(g-i)=0.07$ mag. For the red GC subpopulation, we find $\mu (g-i)=1.08$ mag, with a dispersion of $\sigma(g-i)=0.12$ mag. For comparison, \citet{Faifer11} found the following values based on Gemini/GMOS photometry: $\mu (g-i)=0.78$, $\sigma(g-i)=0.09$, and $\mu (g-i)=1.08$, $\sigma(g-i)=0.11$, for the blue and red subpopulations, respectively.
The local minimum of the combined double-Gaussian is at $(g-i)=1.0$ mag. This is the value adopted to separate blue and red GCs in our ground-based imaging.

\section{Photometric results}
\label{sec:photresults}

We now use our photometrically-selected GCs to study two important relations: how the GC number density and  the GC colours vary with radius. The first is an essential observable for dynamical modelling of galaxies, but it is also needed to quantify contamination from interlopers. The shape of the second relation is an important prediction of theories of the hierarchical growth of galaxies. 
   
\subsection{Globular cluster surface density}

We divide our sample into blue and red GCs at $(g-i)=1.0$ mag. We count the number of GCs in elliptical annuli centered on NGC~4649 and we divide this number by the area of the annulus. Saturated stars, surrounding galaxies and chip gaps in the CHFT images were masked out. For each bin, we compute the Poissonian error using $\sqrt{N} / \mbox{ Area}$, where $N$ is the number of objects per bin. 

As our ground-based imaging is incomplete near the galaxy centre, we supplement our data with the ACS GC surface density profile from \citet{Mineo14} (their Figure 2), which covers the innermost 5 arcmin of the galaxy. This was obtained selecting GCs brighter than $z<22.2$ mag and it was renormalized to match the galaxy surface brightness. Therefore, the ACS surface density profile does not represent absolute counts of objects with $z<22.2$. 
We construct the ground based surface density profile selecting GCs about 1 magnitude fainter with respect to ACS. To account for this normalization difference, we count the GCs selected with our criteria in the radial range $150<R<270$ arcsec, and we compare it to the number of GCs from \citet{Mineo14} (their Figure 2) in the same radial range. We find that a factor of 4.3 and 5.7 is needed to match the ACS-based surface density profile for the blue and red GCs respectively, with our ground-based surface density profiles. 
 
\begin{figure}
\centering
\includegraphics[width=\columnwidth]{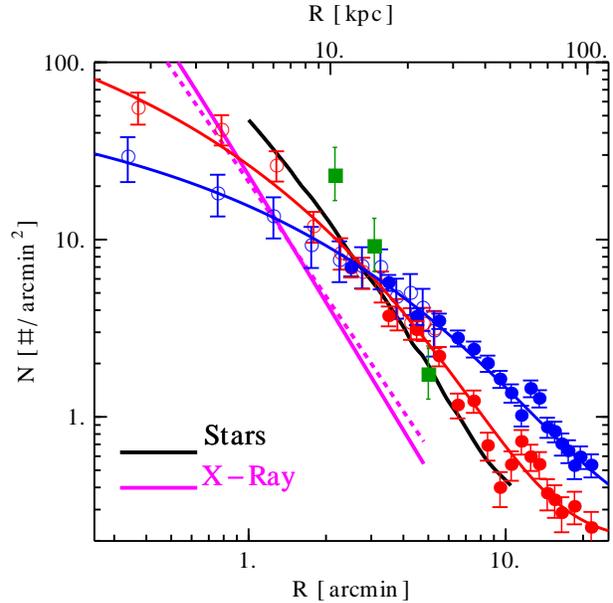} 
\caption{Globular cluster surface density profiles. Blue and red GCs are colour-coded accordingly. Open and filled circles are from \textit{HST}/ACS and from ground-based photometry, respectively. The two solid lines are the best fits to equation \ref{eq:NGC}. Green squares are PNe data. The black line is the arbitrarily re-normalized stellar surface brightness profile of NGC~4649 in the $V$-band from \citet{Kormendy09}. The magenta thick and dashed lines are the outer slope of the X-ray surface brightness profile from \textit{ROSAT} and from \textit{Chandra}, respectively. The red GCs are more centrally concentrated than the blue GCs.}
\label{fig:sd}
\end{figure}

The combined GC surface density profiles for the blue and red GCs are shown in Figure \ref{fig:sd}.

At first glance, our results show the typical spatial features seen in other GC systems: the red GCs are more centrally concentrated, while the blue GCs are more extended \citep[e.g.,][]{Bassino}.  
The flattening of the number density at large radii is due to contaminants (faint Galactic stars and unresolved galaxies), and marks the detectable boundary of the GC system. To quantify the background level $bg$, assumed to be constant across the entire image, we fit the blue and red GC data with a modified \citet{Sersic}:
\begin{equation}
N_{\rm GC}(R) = N_e \times \exp \left( -b_n \left[ \left( \frac{R}{R_e} \right)^{1/n} -1 \right] \right) + bg,
\label{eq:NGC}
\end{equation}
where $n$ is the S\'ersic index and $N_e$ is the surface density at the effective radius $R_e$. The parameter $b_n$ is linked to $n$ via $b_n = 1.9992n - 0.3271$. 

The best fits to equation \ref{eq:NGC} are shown in Figure~\ref{fig:sd}. For the red GCs, we find $R_e=4.1\pm1.1$ arcmin $=20\pm7$ kpc, $n=2.5\pm0.5$ and $bg=0.2\pm0.1$ objects$/$arcmin$^2$. For the blue GCs, we find $R_e=12.6\pm7.1$ arcmin $=60\pm34$ kpc, $n=2.4\pm0.6$ and $bg=0.23\pm0.19$ objects$/$arcmin$^2$. Therefore the blue GC system is a factor of three more extended than the red GCs.

In Figure \ref{fig:sd}, we also compare the GC spatial distribution with the surface density of the following tracers:
\begin{inparaenum}
\item the stellar surface brightness profile of NGC~4649 in the $V$-band from \citet{Kormendy09}; 
\item the X-ray surface brightness profile from \textit{ROSAT} \citep{OSullivan} and from \textit{Chandra} \citep{Paggi14}. In this case, we show a $\beta$-model best fit to X-ray data. The best fit is shown up to the radius where the X-ray emission drops below the background level ($\sim 4.8$ arcmin) for the \textit{ROSAT} data.
\item Planetary nebulae (PNe) surface density data from \citet{Das11}, with the caveat that these were obtained selecting objects within a cone aligned with the galaxy major axis.
\end{inparaenum}
The three tracers were arbitrarily re-normalized in order to be compared with the GC surface density profiles. 

We can see that the slope of the stellar and X-ray surface brightness is qualitatively similar to the slope of the red GCs, with the caveat that this similarity is dependent on the arbitrary normalization. X-ray haloes are usually more extended than the stellar component \citep{Sarazin01,Forte05,Forbes12}. Results from \textit{Chandra} and from \textit{ROSAT} are consistent with each other. 
The similarity between stars--PNe and red GCs is expected if the galaxy bulge and the red GCs formed in a similar fashion at similar epochs \citep[e.g.,][]{Shapiro10}. 

We follow the approach of \citet{Forbes12} to quantify the similarities between the distribution of GCs and X-ray gas. We fit the GC density profiles in Figure \ref{fig:sd} with a $\beta$-model, which is a cored-power law described by $I(r) \propto r^{-3 \beta + 0.5}$, where $\beta$ is the power-law slope we are interested in. The background values $bg$ derived above were taken into account during the fit. A fit to the \textit{ROSAT} data returns $\beta_X=0.56\pm0.08$ \citep{OSullivan}. Using new deep $Chandra$ data from \citet{Paggi14}, one obtains consistent results with $\beta_X=0.52\pm0.01$. 
For the red and blue GCs, we find $\beta_R=0.48 \pm 0.03$ and $\beta_B=0.44 \pm 0.05$, respectively. In this case the larger uncertainty reflects the uncertainty on the GC background value. This exercise shows the outer slope of blue and red GCs are consistent within the errors, but the red GC profile is slightly steeper and more similar to the X-ray slope.

\begin{figure}
\centering
\includegraphics[width=\columnwidth]{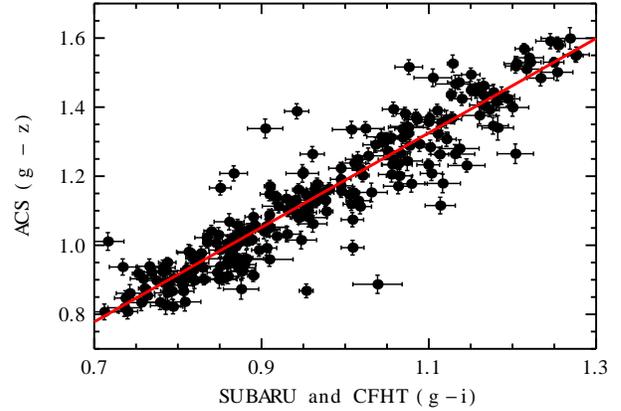} 
\caption{Photometric calibration between ground-based and space-based observations. Data points represent partially resolved GCs in \textit{HST} images, with constraints on galactocentric distance and magnitude (see text). The red line is a weighted least-squares fit to the data.}
\label{fig:colorcalibration}
\end{figure}

\begin{figure}
\centering
\includegraphics[width=\columnwidth]{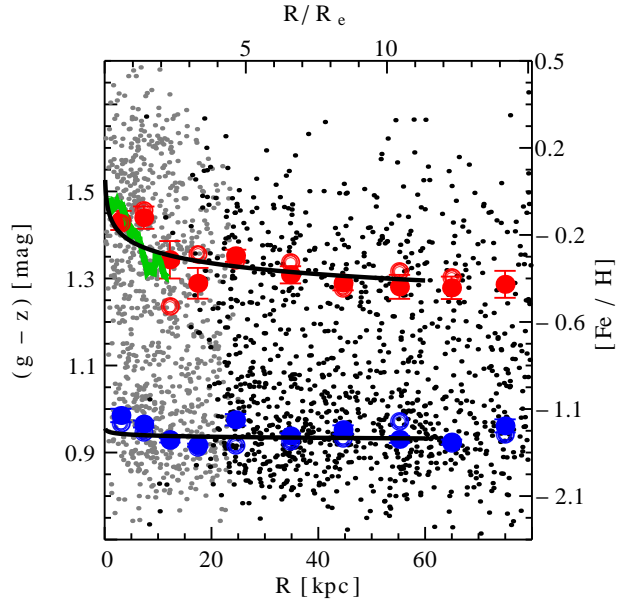} 
\caption{Radial variation of GC colours. Grey and black small points are GC candidates from \textit{HST} observations and from our ground based imaging, respectively. Black lines are best fits to the data up to 60 kpc (see text). Blue and red filled large circles show the colour peaks from KMM for a given radial range. Blue and red open large circles represent the peaks of the Gaussian-smoothed colour histogram in each bin (see text). The green line is the stellar metallicity profile converted to $(g-z)$ (see text). The GC subpopulation colour gradients are steeper towards the galaxy centre.}
\label{fig:gradients}
\end{figure}

\subsection{Globular cluster colour gradients}
\label{sec:gradients}

We study the radial variations of GC colours by combining \textit{HST} and ground-based imaging. It is more efficient to calibrate both datasets to a common photometric colour and then merge them. We choose to transform the ground-based $(g-i)$ to ACS $(g-z)$ because this can be then be converted to metallicity through the conversion of \citet{Peng06}.

As in the previous section, we match the objects in common between \textit{HST} and the combined Subaru$+$CFHT catalogue. We require ACS sources to be partially resolved with sizes $r_h < 10$ pc and magnitudes $19<i<22$ mag. We only consider sources with galactocentric radius $100<R<400$ arcsec because the ground-based imaging is unreliable within 100 arcsec. 

We find 241 objects obeying the above criteria. A weighted last-squares fit to the data returns the colour transformation we are interested in:
\begin{equation}
(g-z) = (-0.18\pm0.03) + (1.36\pm0.03) (g - i)
\label{eq:conversion}
\end{equation}
The best fit, along with the data, is shown in Figure \ref{fig:colorcalibration}. The rms of the data with respect to the best fit line is 0.08 mag. The quality of the fit is satisfactory. We also convert the ACS $g$ magnitudes to ground based $g$ magnitudes. We find $g_{\rm ACS} = (0.62\pm0.20) + (0.97\pm0.01) g$ with an rms scatter of 0.09 mag. Similarly, we find $z_{\rm ACS} = (0.07\pm0.25) + (0.99\pm0.01) i_{\rm ACS}$, with a scatter of 0.13 mag. We apply eq. \ref{eq:conversion} to our ground-based GC catalogue and we merge this into the ACS catalogue. Repeated objects are counted once and priority is given to objects with ACS photometry. 

We construct the colour gradients by selecting $(21<z<23)$ (consistently with \citealt{Strader13}) to avoid contamination from interlopers at faint magnitudes and from the blue-tilt at bright magnitudes. We also clip objects within 5 arcmin of large galaxies in the field to avoid mixing their GC systems with that of NGC~4649. The results are shown in Figure \ref{fig:gradients}.

We can see that the ACS and ground-based datasets agree and that they are complementary to each other. We study the colour profiles out to $\sim 16$ arcmin $\sim80$ kpc, as dictated by the background estimate from equation \ref{eq:NGC}. This radius corresponds to 15 $R_e$, where $R_e=66$ arcsec is the effective radius of the diffuse starlight. 

We compute GC colour radial profiles using two different methods. First, we bin the data radially with a 5 kpc bin size for the innermost 20 kpc, and with a 10 kpc bin size for $R>20$ kpc. For each bin, we fit a double Gaussian with KMM, and derive the blue and red colour peaks of the fitted Gaussians. 
Second, we smooth the colour distribution with a Gaussian kernel of 0.05 mag and we find the blue and red peaks of the resulting smooth distributions. This second method is model independent, whereas the first method requires assumptions that may not provide a realistic representation of the data. 
The results are shown in Figure \ref{fig:gradients} as filled points for the first method (KMM) and open points for the second method (smooth kernel), respectively. We also compare our results with the stellar metallicity profile from \citet{Pastorello14}, converted from $[Z/H]$ to $(g-z)$ using equation B1 from \citet{Usher12}.

We can see that the results in Figure \ref{fig:gradients} are not very sensitive to the method adopted to compute the colour peaks. Both the blue and the red GCs show negative gradients within $\approx 20 $ kpc, as already found in \citet{Strader12}. Outside this radius, the colour gradients are consistent with being constant. The overall slope of the stellar colour profile is in good agreement with the results for the red GCs. This provides further evidence that stars from the galaxy spheroid and red GCs were assembled in a similar fashion, at least in the very central regions.

We quantify the gradients by fitting the function $(g-z)~=~a \log(R/R_e) + b$, where $R_e=5.3$ kpc is the stellar effective radius of NGC~4649. We fit all individual GC datapoints in Figure \ref{fig:gradients} up to 60 kpc to minimize contamination from background sources. We divide the two GC subpopulations by adopting a cut at $(g-z)>1.2$ for the red GCs and $(g-z)<1.1$ for the blue GCs, respectively. This is done to avoid mixing between the two GC subpopulations at intermediate colours. For the red GCs, we find $a=-0.07\pm0.01$ and $b=1.44\pm0.01$. For the blue GCs, we find $a=-0.005\pm0.010$ and $b=0.94\pm0.01$, indicating an overall shallower colour gradient with respect to the red GCs.

The red GC colour gradient is wiggly in the range between 10 and 20 kpc, suggesting the presence of substructures in the data \citep{Strader12,DAbrusco14}, and it is very steep inside 10 kpc. This is due to a group of GCs with very red colours $(g-z)>1.5$, which populate only the innermost 10 kpc of the galaxy.

\begin{figure*}
\centering
\includegraphics[scale=0.5]{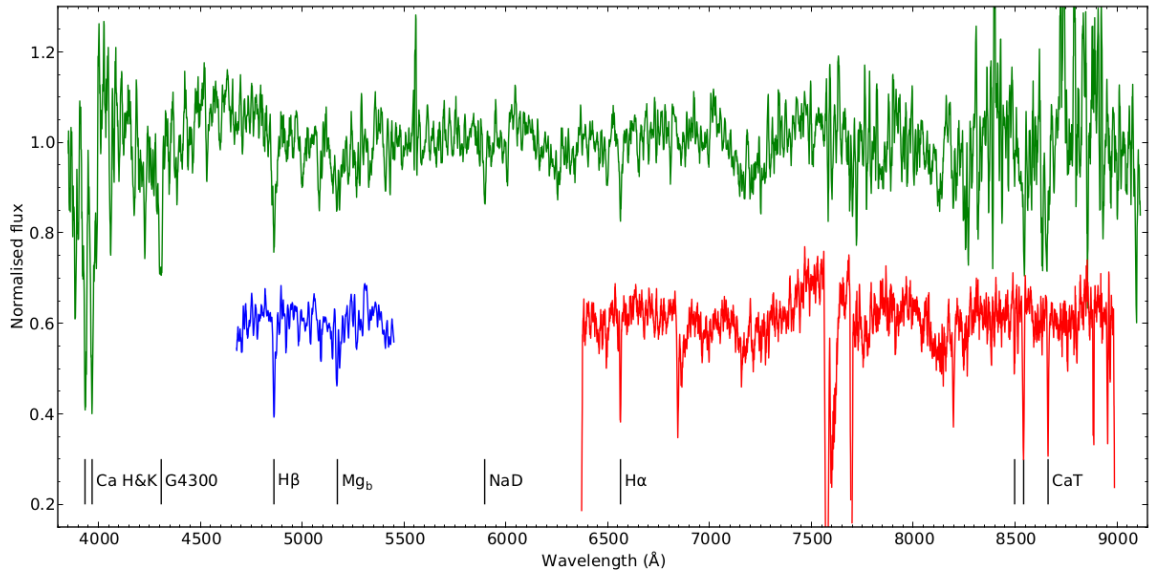} 
\caption{Spectrum of a globular cluster of NGC~4649 as seen from the three multi-object spectrographs used in this paper. Red is DEIMOS, blue is GMOS and green is Hectospec. Spectra are shifted to the rest frame wavelength. An arbitrary vertical offset is applied. Important absorption lines are labelled. Note that the Hectospec spectrum was corrected for telluric absorption, while the DEIMOS spectrum was not, which explains why the strong water absorption feature at $\sim$7700\AA\ appears in the latter but not the former. The globular cluster's ID is D28. Its apparent magnitude is $z=20.3$, and its radial velocity is $1160\pm7$\kms .}
\label{fig:spectrum}
\end{figure*}

\section{Spectroscopy}
\label{sec:spectroscopy}

Our spectroscopic sample comes from three multi-object spectrographs: Keck/DEIMOS, MMT/Hectospec and Gemini/GMOS. The footprint of the observations of each instrument is shown in Figure~\ref{fig:imag}. The spectral coverage of the three spectrographs is diverse, but some spectral regions overlap. Figure \ref{fig:spectrum} shows one of the nine GC spectra observed by all three spectrographs. In this section we describe the acquisition and reduction of the spectroscopic data.

\subsection{Keck/DEIMOS}
\label{sec:DEIMOS_spec}

DEIMOS is a multi-object spectrograph mounted on the 10 meter Keck II telescope \citep{Faber03}. One DEIMOS mask covers 15$\times$6 arcmin$^2$ on the sky and it can contain up to 150 slits. 

A total of four DEIMOS masks were observed around NGC~4649.
Three masks were observed on 2012 Jan 17, and the fourth mask was observed on 2013 Jan 11. Slitmasks were designed giving priority to ACS GC candidates with size measurements. If no space-based photometry was available, slits were assigned to bright sample of GCs selected from the CFHT imaging discussed in Section \ref{sec:imaginganalysis}. Suprime-Cam images were not yet available at the time of the mask design.
 
All observations were performed in sub-arcsec seeing conditions. The exposure time was 2 hours per mask. The instrument setup consisted of a 1200 l/mm grating, a dispersion of 0.33 \AA/ pixel, a slit width of 1 arcsec and the OG550 filter.
With this setup the GC spectra cover the region from the H$\alpha$ line (6563 \AA)  to the Calcium Triplet (CaT) feature (8500-8700 \AA), with a median velocity accuracy of $\sim$ 15 \kms.
The spectra were reduced with a dedicated pipeline, which returns calibrated, sky-subtracted one-dimensional spectra \citep{Cooper12,Newman13}. 

GC radial velocities were measured with iraf/fxcor using thirteen stars of different spectral type as templates observed using the same set-up. The average velocity scatter between different stellar templates is $\sim 3\kms$ for all four DEIMOS masks.
The Doppler shift was computed solely from of the CaT lines, because the H$\alpha$ line was not always covered with our instrument setup. The final GC radial velocity was taken as the average of the values from the thirteen stellar templates. The redshift-corrected spectra were inspected by eye to check that the CaT lines laid at the rest frame wavelength. The error budget on the velocity takes into account both the internal error given by fxcor and the scatter due to different stellar templates. 

\subsection{Gemini/GMOS}
\label{sec:GMOS_spec}

GMOS is a multi-object spectrograph and imager mounted on the 8-meter GEMINI telescopes \citep{Hook04}. The GMOS camera has three 2048 $\times$ 2048 pixel CCDs with a scale of 0.07 arcsec/pixel and a field of view 5.5 $\times$ 5.5 arcmin$^2$. GMOS was used to obtain spectra for globular clusters in five fields in NGC 4649 (see Figure\ref{fig:imag}). GC selection was performed as described in \citet{Faifer11}. GC spectroscopy for a field positioned near the galaxy center (Field 1) was obtained in May 2003 on Gemini North and was presented in \citet{Bridges06} and \citet{Pierce06}, while spectroscopy for four additional fields (Fields 2-5) was obtained in the spring of 2010 on Gemini South and presented in this paper. 

GC candidates for follow-up spectroscopy were selected using Sloan \textit{g$^\prime$} and \textit{i$^\prime$} images taken with GMOS-N in April 2002. Images were combined and median-filtered and run through iraf/daophot to obtain photometry for point-like objects. The instrumental photometry was calibrated using F555W and F814W \textit{HST}/WFPC2 photometry. A final sample of globular cluster candidates was obtained using magnitude (9 $< V <$ 22.5 mag) and colour (0.75 mag $<$ $V-I$ $<$ 1.4 mag) cuts. Extended sources were eliminated by eye, leaving 250 globular cluster candidates. A GMOS multi-slit mask was made for Field 1 using the standard Gemini GMMPS software, containing 39 object slits. 

For the Fields 2-3-4-5 (program ID GN-2010A-Q-37), GMOS images in the SDSS $g^\prime r^\prime i^\prime$ filters were taken in 2007A and 2009A to select GC candidates for multi-slit spectroscopy. Colour and magnitude cuts and image classification using SExtractor were used to obtain a final sample of GC candidates in each of the four fields. The GMMPS software was used to create multi-slit masks, and with the use of filters to restrict wavelength coverage, we were able to put slits on between 48 to 69 globular cluster candidates in the four fields. Some marginal objects were included to fill the mask, and we also placed some slits on the diffuse galaxy light. 

Spectroscopy for Field 1 (program ID GN-2003A-Q-22) was obtained using GMOS-N in multi-slit mode in May and June 2003. A total of 8 $\times$ 1800 sec exposures were obtained with a central wavelength of 5000 \AA, and 8 $\times$ 1800 sec were obtained with a central wavelength of 5050 \AA, giving a total of 8 hours on-source time. 
We used the B600 grating, giving a dispersion of 0.45 \AA/pixel, a resolution of $\sim$ 5.5 \AA, which corresponds to 320 $\kms$ velocity resolution. Spectral coverage was typically from 3300$-$5900 \AA, and the seeing ranged from 0.65 to 0.9 \arcsec over the four nights of observation. Bias frames, dome flat-fields and CuAr arcs were also taken for calibration. 

Spectroscopy for the remaining four fields was obtained at GMOS-S in the spring of 2010. 
We obtained between 5 to 7.5 hours on-source per field, and conditions were good with seeing between 0.5 and 1.0 arcsec. 

Data reduction was carried out using the Gemini/GMOS \textsc{iraf} package, and consisted of the following steps: (i) a final spectroscopic flat-field was created; (ii) object frames were bias-subtracted and flat-fielded; (iii) the arc frame was reduced and used to establish the wavelength calibration; (iv) object frames were wavelength calibrated and sky-subtracted; (vi) 1D spectra were extracted from each object frame and median combined. 

Iraf/fxcor was used to obtain radial velocities for each object via cross-correlation, using the MILES spectral library \citep{Sanchez06} as template. In this case, the template with the highest cross-correlation coefficient was used, although the relative scatter between templates was $\sim 4-5\kms$, well within the measurement errors.

Before accounting for overlaps, there were $\sim$ 185 objects with reliable velocities in fields 2-3-4-5. Astrometry and photometry for this sample were obtained from \citet{Faifer11}. Globular clusters from Field 1 \citep{Bridges06} were then added to this list, and objects observed in more than one field were identified. 

\subsection{MMT/Hectospec}
\label{sec:Hectospec_spec}

Hectospec is a multi-fibre spectrograph mounted on the 6.5 meter MMT telescope \citep{Fabricant05}. It has a 0.78 deg$^2$ field-of-view which can be filled with up to 300 fibers.

GC candidates were selected from the MegaCam and ACS images. Bright objects at large radii were favoured, although this naturally increases the contamination from Galactic stars. Due to the wide Hectospec field-of-view, GC systems in galaxies surrounding NGC~4649 (like NGC~4621 and NGC~4638) were also observed. Therefore, some contamination is expected from these galaxies.

One single Hectospec field was observed on May 16 and May 19, 2012, using a 270 l/mm grating, with a dispersion of 1.21 \AA pixel, and a spectral resolution of $\sim 5$ \AA, which corresponds to a velocity resolution of $\sim 300 \kms$.

Raw data were reduced as described in \citet{Mink07}. Heliocentric radial velocities were measured by cross-correlating the science spectra with a Hectospec template of an M31 GC. Errors were estimated through Monte Carlo simulations with an additional 8 \kms to account for the wavelength calibration \citep[see also ][]{Strader11}. 

\section{Outcome of the spectroscopic follow-up}
\label{sec:specresults}

We observed $\sim$ 600 objects with DEIMOS, targeting $\sim 500$ photometrically selected GCs. We found 304 spectra (including duplicates between masks) consistent with the systemic velocity of NGC~4649. 23 additional GCs were discarded as ``marginal'' because the team could not reach a consensus on whether or not the features seen in their spectra were actual spectral lines. In addition, we found 13 Galactic stars and 56 background galaxies. The latter were identified via the strong emission lines in their spectra. However we do not measure the redshift of background galaxies. The remaining slits were unclassified, because the signal-to-noise was too low to retrieve any physical information from the spectrum. After accounting for repeated objects, the median velocity uncertainty of the DEIMOS GC sample is $\Delta v = 11 \kms$.

We observed a total of 187 objects with Gemini/GMOS, finding 172 extragalactic GCs, 14 Galactic stars and 1 background galaxy. The 172 GCs also include the 38 GCs observed by \citet{Pierce4649}. GMOS maps the innermost 6 arcmin of NGC~4649 and it overlaps nicely with the DEIMOS field-of-view. The median velocity uncertainty of this sample is $\Delta v = 12 \kms$.

We observed 478 objects with Hectospec, and we found 102 extragalactic GCs (including duplicates), 157 background galaxies and 119 Galactic stars. The remaining fibres were unclassified. The median velocity uncertainty of this sample is $\Delta v = 14 \kms$, but in Section \ref{sec:repeated} we will explain that the effective uncertainty is larger than this value.

\section{Repeated measurements}
\label{sec:repeated}

We quantify how the radial velocity of a given GC compares to the radial velocity of the same GC observed over multiple spectrographs. We denote radial velocities from DEIMOS, GMOS and Hectospec wih $v_{\rm D}$, $v_{\rm G}$ and $v_{\rm H}$ respectively. The uncertainties of the repeated objects are the sum in quadrature of the uncertainties of the single objects. The comparison between the three catalogues is illustrated in Figure \ref{fig:repeated} and discussed in the followings. 

\subsection{Duplicates from the same spectrograph}

We find 26 and 33 GCs repeated across different DEIMOS and GMOS masks, respectively. 
For DEIMOS, the weighted-mean of the velocity difference of the objects in common is $\langle v_{\rm D} - v_{\rm D} \rangle = -0.6 \kms$, with an root-mean-square difference rms$= 10 \kms$. We conclude that the DEIMOS radial velocities are robust, and no additional velocity offset is needed. 

For GMOS, we find field to field offsets in the range $| \langle v_{\rm G} - v_{\rm G} \rangle | = 25 - 40 \kms$, which probably reflects the different instrument set-ups of the five GMOS fields. The repeated radial velocity measurements from the five fields are generally consistent within 2$\sigma$. We decided to simply average the repeated GMOS GCs with no rigid offset, and use the comparison with the DEIMOS sample to renormalize the velocity uncertainties of the GMOS sample (see \S \ref{sec:calibration_velocities}). 

\subsection{Duplicates from different spectrographs}

There exist 67 GCs in common between the DEIMOS and GMOS catalogues. This sample has $\langle v_{\rm D} - v_{\rm G} \rangle = -0.6 \kms$, and the root-mean-square is rms$= 23 \kms$. 

The DEIMOS and Hectospec samples have 21 GCs in common. The average velocity difference is $\langle v_{\rm D} - v_{\rm H} \rangle = 9.3 \kms$, with a scatter of rms$=30 \kms$. 

\begin{figure}
\centering
\includegraphics[width=\columnwidth]{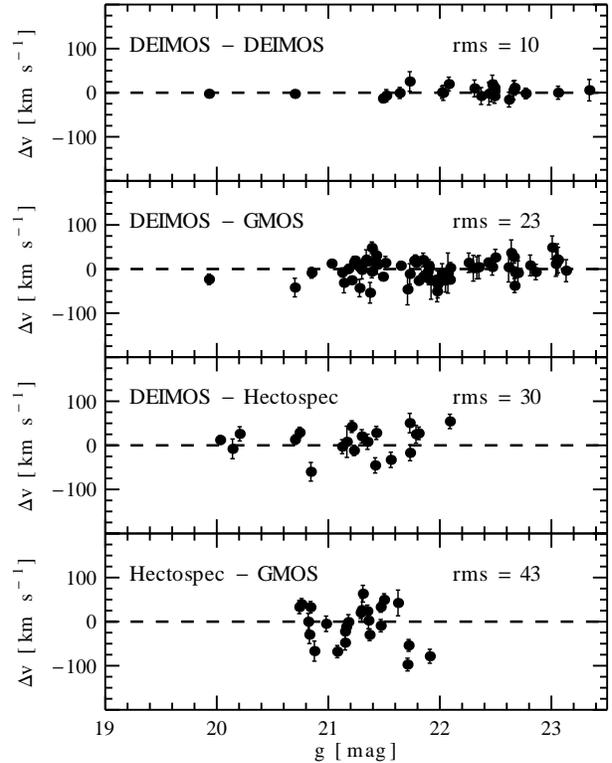} 
\caption{Repeated radial velocity measurements. Each panel shows the velocity difference for the objects in common between different catalogues (labelled on the top left) vs the magnitude in the $g$ band. The top panel shows the comparison between GCs observed over different DEIMOS masks. The rms scatter in units of $\kms$ of the velocity difference is reported on the top right of each panel. The velocity agreement is usually good (always $<100\kms$), although the velocity uncertainties from GMOS and Hectospec are probably underestimated (see \S \ref{sec:calibration_velocities}).}
\label{fig:repeated}
\end{figure}

Lastly, we compare the GMOS and Hectospec catalogues. In this case the average velocity difference is $\langle v_{\rm H} - v_{\rm D} \rangle = -6 \kms$, with rms$=43 \kms$. In the bottom panel in Figure \ref{fig:repeated} it is clear that some of the 25 objects in common between these two catalogues, scatter up to 5$\sigma$ from each other. 

\subsection{Calibration of velocity uncertainties}
\label{sec:calibration_velocities}

By comparing radial velocities from different instruments we have seen that the average velocity difference is generally small ($< 15 \kms$), comparable to our DEIMOS uncertainties. However, the large scatter in $\langle \Delta v \rangle$ compared to the small error bars of the single datapoints, implies that the velocity uncertainties of GMOS and Hectospec catalogues are probably underestimated with respect to those of DEIMOS. Therefore, we need to re-normalize the velocity uncertainties of the first two spectrographs to those of DEIMOS.

We follow the method of \citet{Strader11}, and we multiply the uncertainties of the GMOS and Hectospec by a factor $\tau$ until the $\chi^2$ difference between the velocities of the two catalogues is $\chi^2 \sim 1$. This normalization counts as a first-order correction to the velocity inconsistencies discussed above. We find that GMOS uncertainties need to be inflated by a factor of $\tau=1.6$ with respect to DEIMOS. Once the uncertainties have been normalized, we weight-average the radial velocities of the GCs in common and we add the unique GMOS GCs to the DEIMOS catalogue. No rigid velocity offset has been applied to the GMOS velocities, as this is negligible ($\langle v_{\rm D} - v_{\rm D} \rangle = -0.6 \kms$). 

Next, we compare the combined DEIMOS$+$GMOS catalogue to Hectospec. We renormalize the uncertainties of the Hectospec GCs by a factor of $\tau=2.4$. Also in this case the velocity difference between the two datasets is negligible $\Delta v=5 \kms$, therefore no offset is applied. We merge Hectospec to the master catalogue following the above procedure. Our master catalogue is composed of 447 spectroscopically confirmed stellar objects (GCs, UCDs and dwarf galaxies). 

\section{Building up the spectroscopic master catalogue}
\label{sec:master}

Before proceeding, we assign photometric measurements (i.e. magnitudes and sizes) to all confirmed objects. We match the position of confirmed GCs with $all$ sources, both point-source and extended, extracted from ground based and HST imaging. We transform the ground based colours of the confirmed GCs from $(g-i)$ to $(g-z)$ using Equation \ref{eq:conversion}. This is more convenient than transforming $(g-z)$ to $(g-i)$ because 60 per cent of the confirmed GCs have ACS photometry. 

\subsection{Clipping GCs from neighbour galaxies}
\label{sec:GCcont}

The large field-of-view of Hectospec allows us to confirm GCs in some galaxies near NGC~4649 (see Figure \ref{fig:imag}). These GCs are treated as interlopers because they do not follow the gravitational potential of NGC~4649. The galaxies which may contaminate our sample are (from \citealt{Cappellari11}): NGC~4621 $(v_{\rm sys} = 467 \kms)$, NGC~4638 $(v_{\rm sys} = 1152 \kms)$, NGC~4647 ($v_{\rm sys}=1409\kms$, from NED), NGC~4660 ($v_{\rm sys}=1087\kms$). We remind the reader that NGC~4649 has $v_{\rm sys} = 1110 \kms$ \citep{Cappellari11}.

Of the 7 GCs around NGC~4621, five are probably associated with this galaxy because their radial velocities are very similar to that of this galaxy as we will show in Section \ref{sec:overview}. The remaining two GCs (those in the NGC~4649 direction) have radial velocities consistent with NGC~4649 and they probably belong to this galaxy. Using geometrical criteria, we find 4 GCs associated with the elliptical galaxy NGC~4660. 

We flag one GC as a member of NGC~4638 because it is very close to the center of this galaxy. However, the membership of this GC is impossible to confirm because the systemic velocity of NGC~4638 is almost identical to that NGC~4649.

GCs belonging to the spiral galaxy NGC~4647 (2.6 arcmin from NGC~4649) may also contaminate our GC sample. However, \citet{Strader12} estimated that only $\sim 10$ objects in their GC catalogue are actually associated with NGC~4647. From our spectroscopic GC catalogue, we exclude one GC because of its proximity to NGC~4647 and because the spiral arms compromise the photometry of this object.

\subsection{Clipping surrounding galaxies}
\label{sec:galcont}

In addition to GCs, we also measured radial velocities for some dwarf galaxies in the field. 

With DEIMOS and GMOS, we observed the dE VCC~1982 $(v=945\pm4\kms)$ and the nucleated dE VCC~1963 $(v=1090\pm13\kms)$. For comparison, \citet{Lee4649} found velocities of $976\pm 42 \kms$ and $1027\pm 53\kms$, respectively. We also obtained a radial velocity for a very low surface brightness nucleated dEs (ID 81 from \citealt{Sabatini05}). We found its radial velocity to be $v=1186\pm22\kms$. 

With Hectospec, we observed two dwarf galaxies present in the photometric catalogue of \citet{Grant05}. The first, is the nucleated dE VCC~1951, for which we find $v=1089\pm32\kms$, consistent with the measurement from SDSS DR10 $v=1106\pm13\kms$. The second, is the dE VCC~1986, for which we measure the redshift for the first time, finding $v=850\pm37\kms$.

The object M60-UCD1 is also treated as a galaxy and excluded from our sample \citep{Strader13}.

One GC is only 7 arcsec away from a faint dwarf galaxy without spectroscopic confirmation, but we assume this is a projection effect and that these two objects are not physically connected. The same assumption was adopted for another GC which is $\sim 8$ arcsec away from the centre of a small galaxy without a redshift measurement (J124334.56$+$112727.3 from SDSS DR8).

\subsection{Additional GC datasets}
\label{sec:additional}

We searched the SDSS archive for point sources with spectroscopic confirmation within 1 deg from NGC~4649. We found one unique object, which we added to our master catalogue, consistent with being a GC associated with NGC~4649 (ID SDSSJ124349.56$+$113810.3). We found four other GCs with SDSS spectroscopic confirmation, but these were not added to our catalogue because they already observed and confirmed with Hectospec.

\begin{figure}
\centering
\includegraphics[width=\columnwidth]{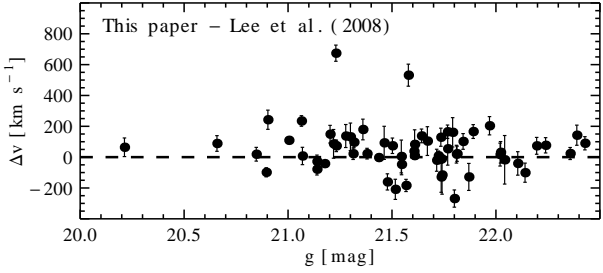} 
\caption{Comparison with the data of \citet{Lee4649}. The plot is the same as Figure \ref{fig:repeated}, but for our combined spectroscopic master catalogue versus the CFHT/MOS GC dataset of \citet{Lee4649}. Two objects scatter up to $700\kms$ with respect to the $\Delta v =0$ dashed line.}
\label{fig:Lee}
\end{figure}

\citet{Lee4649} published radial velocities for 93 GCs using CFHT/MOS with a median velocity uncertainty of 50\kms. Their observations explored an area $\approx 14 \times 14$ arcmin$^2$, which overlaps the spectroscopic observations from our paper. In fact, we found 61 GCs in common between our spectroscopic master catalogue and Lee et al.'s dataset. The median velocity difference is 38 $\kms$, with an rms of rms=$156\kms$. A more careful inspection of the dataset, shown in Figure \ref{fig:Lee}, reveals that the large rms value is driven by two objects (ID 201 and ID 226 using Lee et al. nomenclature) which scatter up to $600\kms$ or $(5\sigma)$ from the ideal $\Delta v = 0\kms$ line. Removing these two extreme objects, the velocity difference rms decreases to $113\kms$, still a factor of two larger than measured in our dataset.
 
The magnitude of this effect is comparable to that of the outliers found in M~87 by \citet{Strader11} when comparing to the CFHT/MOS dataset of \citet{Hanes01}. We investigated the reason for this offset. We excluded mismatch because the outliers are not in overcrowded regions. We excluded a low signal-to-noise issue, because the two objects have intermediate magnitudes and because fainter duplicates have radial velocities in reasonable agreement with our dataset. As we could not identify the reason for this disagreement, we decided not to include the Lee et al. dataset in our DEIMOS catalogue because additional unknown outliers amongst the 32 unique Lee et al. GCs might skew our GC kinematic results. 

\section{Overview of the spectroscopic catalogue}
\label{sec:overview}

We compile a list of 448 unique spectroscopically confirmed objects around NGC~4649. After clipping dwarf galaxies and GCs from neighbour galaxies, we are left with 431 unique GCs belonging to NGC~4649. The spectroscopic catalogue is given in Table \ref{tab:table}. In the following, we review different projections of our parameter space, which involves photometric quantities (magnitude and colours) as well as physical quantities, such as the radial velocity and the physical size.

\begin{table*}
\centering
\label{mathmode}
\begin{tabular}{@{}l l l l l l l l l l l l l l l l l}
\hline
ID & ID(S$+12$) & RA & DEC & $(g-z)$ & $z$ & $R$ & $v$ & $\delta v$\\
\hline
 &  & [deg] & [deg] & [mag] & [mag] &  [arcsec] & [km/s] & [km/s] \\
\hline
412 &	-   &	191.09555 &	11.45407 &	1.095 &	20.177 &	694 &	1240 &	32 \\
295 &	-   &	190.85417 &	11.52163 &	0.991 &	20.455 &	237 &	1034 &	23\\
114 &	C4  &	190.94971 &	11.59446 &	1.140 &	19.943 &	197 &	1247 &	22\\
59  &	B12 &	190.92905 &	11.53665 &	1.574 &	19.759 &	75  &	612  &	6\\
339 &	-   &	191.04245 &	11.57867 &	0.948 &	18.483 &	420 &	1450 &	28\\
353 &	-   &	191.01315 &	11.61024 &	1.331 &	21.138 &	387 &	771 &	14\\
96  &	B87 &	190.97552 &	11.51810 &	1.361 &	21.552 &	233 &	1076 &	21\\
160 &	D28 &	190.90526 &	11.52134 &	0.948 &	20.352 &	129 &	1160 &	7\\
333 &	-   &	190.80938 &	11.57225 &	1.529 &	21.760 &	354 &	1424 &	8\\
295 &	-   &	190.85417 &	11.52163 &	0.991 &	20.455 &	237 &	1034 &	23\\
...	& ...	& ...	& ...& ...	& ...& ...	& ...& ... \\
\hline
\end{tabular}
\caption{Catalogue of the 431 spectroscopically confirmed GCs associated with NCG~4649. ID(S$+12$) is the ID of the GCs also present in \citet{Strader12}. Right ascension and declination are given in degrees. $(g -z)$ and $z$ are either from Strader et al., or have been converted from $gri$ magnitudes as described in \S \ref{sec:gradients}. $R$ is the galactocentric radius. $v$ is the observed radial velocity of the objects. The quoted uncertainties on $v$, $\delta v$, have been re-normalized to the DEIMOS uncertainties (see text). The full table is available online. }
\label{tab:table} 
\end{table*}

\subsection{Spatial distribution}

\begin{figure}
\centering
\includegraphics[width=\columnwidth]{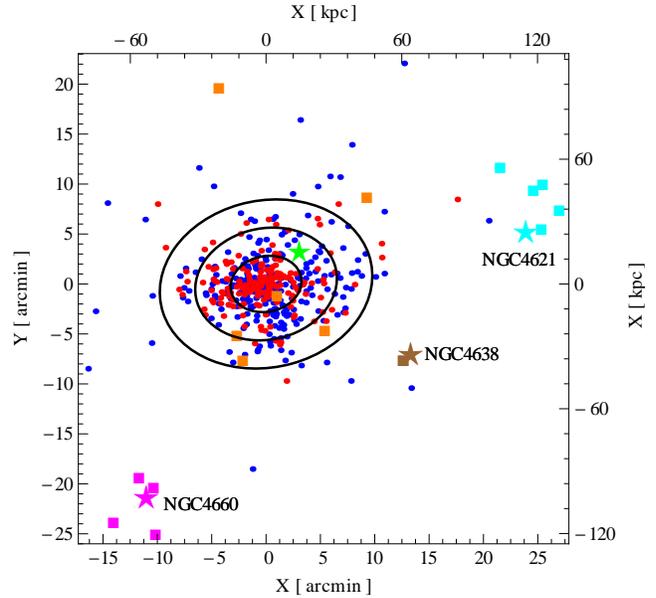} 
\caption{Spatial distribution of the spectroscopically confirmed objects. NGC~4649 is at $(0,0)$. Top and right axes show the co-moving physical scale at the distance of NGC~4649. Confirmed GCs are divided into blue and red subpopulations and coloured accordingly. The green star marks the position of NGC~4647. Orange boxes are dwarf galaxies discussed in the text. Ellipses are schematic isophotes at $3,6,9$ $R_e$. Position of some of the surrounding galaxies and their confirmed GCs are marked with coloured stars and squares, respectively. North is up, east is left.}
\label{fig:pos}
\end{figure}

We start by looking at the positions on the sky of all confirmed objects (Figure \ref{fig:pos}). We confirm GCs up to 100 kpc from NGC~4649. DEIMOS and GMOS contribute to the GCs in the innermost 30 kpc, whereas all objects outside this radius are from Hectospec. In the outermost regions, mostly blue GCs are confirmed as expected from the shape of the GC surface density profile. 

From Figure \ref{fig:pos} is also clear that the GC systems of nearby galaxies are spatially separated from NGC~4649. Additional contamination from these galaxies is unlikely. The paucity of data in the immediate North-West of NGC 4649 is due to NGC~4647, which prevents us from detecting any source.

\subsection{Globular cluster velocity distribution}

\begin{figure*}
\centering
\includegraphics[scale=0.8]{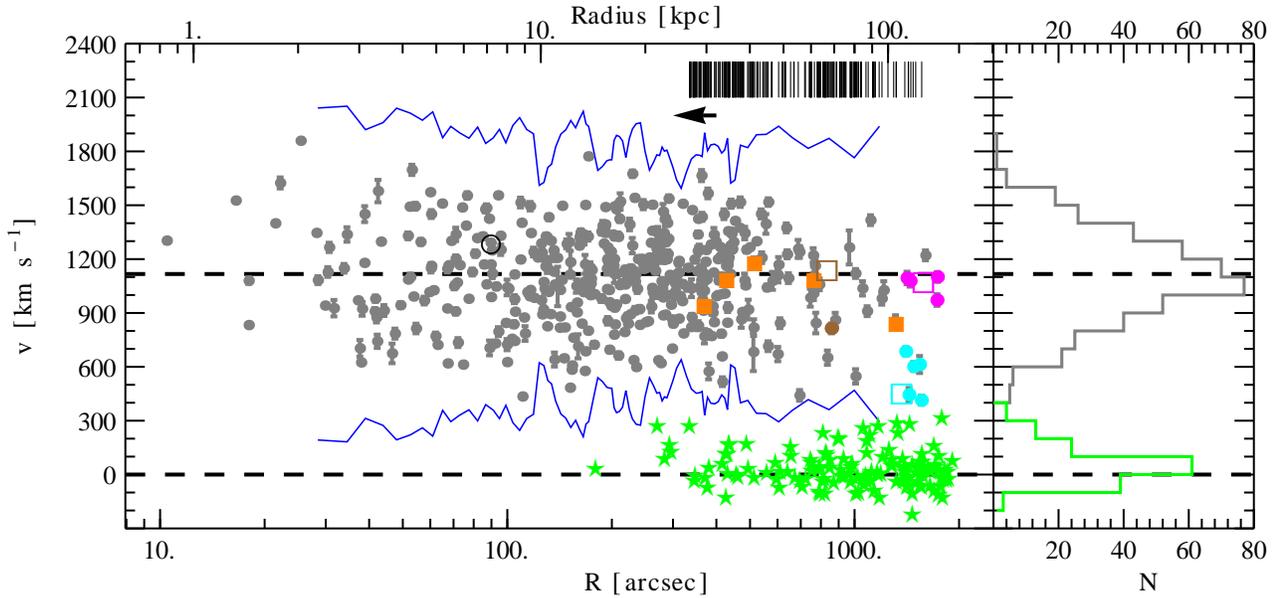} 
\caption{GC velocity distribution vs. galactocentric radius. Shown are all 447 spectroscopically confirmed objects.\textit{ Left panel:} observed radial velocities vs galactocentric distance $R$, shown in physical units on the top axis. Error bars are the velocity measurement errors re-normalized to the DEIMOS uncertainties (see text for details). Grey points are objects associated with NGC~4649; orange squares are the dwarf galaxies discussed in \S \ref{sec:galcont}; the open black circle at $R\sim100$ arcsec is the object M60-UCD1. Cyan points are GCs associated with NGC~4621; magenta points are GCs associated with NGC~4660; the brown point is a GC probably associated to NGC~4638. Open boxes show the position in this diagram of the three galaxies above. Green stars are Galactic stars. The $\pm 3 \sigma$ envelope of the NGC~4649 GCs (grey points only) is shown as a blue lines. Horizontal lines show $v=0\kms$ and $v=v_{\rm sys}=1110\kms$, which is the adopted systemic velocity of NGC~4649. Vertical thin black lines show the position of confirmed background galaxies with arbitrary velocities. \textit{Right panel:} velocity distribution histogram for GCs/UCDs belonging to NGC~4649 only (grey histogram), and Galactic stars (green histogram). Both histograms have the same bin-step of $100 \kms$.  GCs/UCDs are well separated from Galactic stars and background galaxies at all radii.}
\label{fig:RV}
\end{figure*}

In Figure \ref{fig:RV} we plot the GC velocity distribution vs. galactocentric distance. This diagram is useful for identifying outliers which may scatter from the GC system velocity distribution. The latter is shown in the right panel of Figure \ref{fig:RV}. The skewness and kurtosis of the GC velocity distribution (excluding GCs from other galaxies) are $s=-0.05\pm0.11$ and $\kappa - 3 =-0.21\pm0.23$, respectively, suggesting Gaussianity. This is reassuring because we will invoke Gaussianity later on. If we consider only GCs within the innermost 100 arcsec, we estimate the systemic velocity of the GC system to be $v_{\rm sys} =1115 \pm 28 \kms$, very consistent with literature values for NGC~4649. 

We bin the GCs by radius and calculate $\sigma$, which is the standard deviation of the binned velocities with respect to the systemic velocity of the galaxy. This allows us to draw a 3$\sigma$ envelope as a tool to flag GCs deviating more than 3$\sigma$ from the local GC velocity distribution. As can be seen in Figure \ref{fig:RV}, none of the objects shows such a deviation..

The velocity distribution of GCs is distinct from that of Galactic stars and from background galaxies. Therefore, a velocity cut at $v=400\kms$ ensures no contamination from Galactic stars. We note that for $R>700$arcsec, the majority of GCs have blue-shifted radial velocities. Within $R=200$ arcsec, neither Galactic stars nor background galaxies were found thanks to the superior spatial resolution of ACS..
Confirmed GCs from other galaxies are clustered around the systemic velocity of the host galaxy. 

\subsection{Trends with colour}

\begin{figure}
\centering
\includegraphics[scale=.7]{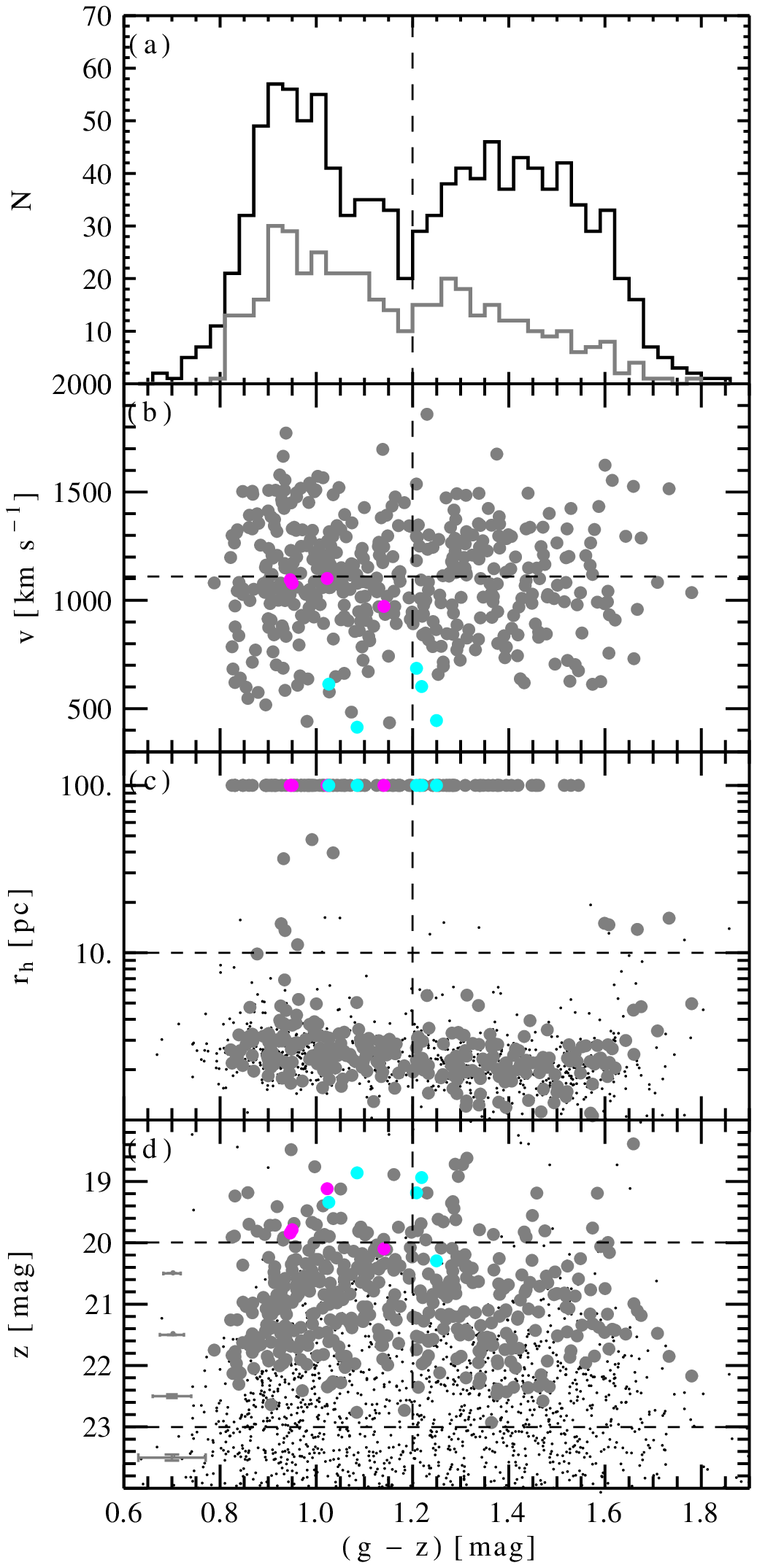} 
\caption{GC trends with colour. Each panel shows the $(g-z)$ colours of the spectroscopically confirmed GCs (large grey points) and of the \textit{HST} GC catalogue (small black points). Coloured points are confirmed GCs from other galaxies as defined in Figure \ref{fig:RV}.  
The vertical line marks the separation between blue (on the left) and red GCs (on the right). 
Objects with ground-based photometry lack size measurements, and they were given $r_h = 100$ pc, as can be seen in Panel \textit{(c)}. The black and grey histograms in Panel \textit{(a)} are the HST dataset and the confirmed GCs, respectively; the horizontal line in Panel \textit{(c)} separates UCDs from GCs; the horizontal lines in Panel \textit{(d)} are the magnitude of $\omega$ Cen at the distance of NGC~4649 and the turn-over-magnitude.}
\label{fig:CE}
\end{figure}

Figure \ref{fig:CE} illustrates how some physical and photometric properties correlate with the observed GC colour. The quantities that we explore are the magnitude, the physical size~$r_h$ and the radial velocity~$v$.

In the top panel we plot the colour histogram of all the \textit{HST} candidates compared to the outcome of our spectroscopic follow-up. The overall colour distribution is well sampled with confirmed GCs. Bimodality is preserved, albeit there may be a deficit of redder GCs. 

The correlation of colours with velocity (Panel \textit{b}) suggests that the blue GCs scatter with respect to $v_{\rm sys}$ more than the red GCs. In other words, the velocity dispersion is larger than that of the red GCs. We will quantify this statement in the next section.

Panel \textit{c} shows how colour correlates with size. The average GC size of the sample is $r_h \approx 2$ pc. We confirm 10 objects with $r_{h} > 10$ pc. We note that the catalogue of \citet{Strader12} lists 22 objects with $r_h > 10$ pc, therefore our return rate is about 50 per cent. We confirm all three objects with $r_{h} > 20$ pc, two of which were already confirmed in \citet{Lee4649}. The paucity of UCDs in this galaxy prevents us from any kinematic analysis of the UCD sample. We define UCDs as objects brighter than $\omega$ Cen and with $r_h > 10$ pc, where we have adopted the size cut adopted by \citet{Brodie11} or \citet{Zhang15}.

It is interesting that very extended objects are either the reddest, or the bluest. Among the blue objects, only 2 can be defined as ``classic'' UCDs, because along with being larger than GCs, they are also bright ($M_z<-11.0$ mag). The remaining blue objects have intermediate magnitudes and are the faint UCDs discussed in \citet{Brodie11}. Regarding the confirmed extended objects with red colours (four in total), these are generally fainter than $z>21$, and never larger than $r_h=20$ pc. These objects are more likely linked to faint-fuzzies \citep{Brodie02}.

Lastly, the bottom panel shows the colour-magnitude diagram of both candidate and confirmed GCs. We notice that some are brighter than $\omega$ Cen (see also Figure \ref{fig:CC}). The M60-UCD1 object has $z=15.6$ mag and is not visible in the plot. The blue-tilt is clearly visible, as well as the fact that the red subpopulation has a larger colour spread compared to the blue subpopulation. We note the presence of $\sim 10$ objects with $z<20$ mag and unusually blue colours $(g-z)<0.95$ mag, which are found outside the ACS field-of-view and therefore lack of size information. It will be interesting to determine why these objects do not tilt towards redder colours like to bulk of blue GCs. We defer such analysis to a future paper.

\section{Kinematics}
\label{sec:kinematics}

So far, we have qualitatively discussed some properties of the spectroscopic sample, with no assumptions about the shape of the GC velocity distribution. This has suggested that blue and red GCs have different properties, as found in other GC systems \citep[e.g.,][]{Schuberth,Pota13}. 
We now quantify these findings and we derive physical quantities capable of characterizing the kinematics of the NGC~4649 GC system. We are interested in its rotation amplitude $v_{\rm rot}$,  velocity dispersion $\sigma$, kinematic position angle $\theta_0$, and root-mean-square velocity dispersion $v_{\rm rms}$.

\subsection{Method}
\label{sec:method}

We start by calculating the root-mean-square velocity dispersion profile $v_{\rm rms}$, which measures the standard deviation of GC radial velocities from the systemic velocity of the galaxy:
\begin{equation}
v_{\rm rms} ^2 = \frac{1}{N} \sum_i ^N (v_i - v_{\rm sys})^2 - (\Delta v_i)^2
\label{eq:vrms}
\end{equation}
where $\Delta v_i$ is the measurement uncertainty of the $i$-th GC. Uncertainties on $v_{\rm rms}$ are obtained using the formula of \citet{Danese}. The $v_{\rm rms}$ can be thought of a measure of the specific kinetic energy of a system $v_{\rm rms} ^2 \approx v_{\rm rot} ^2 + \sigma^2$ and does not require any assumption about the detailed velocity distribution of the sample. However, we are also interested in the relative weights of $v_{\rm rot}$ and $\sigma$ in the final value of $v_{\rm rms}$, which cannot be deduced from equation \ref{eq:vrms}.

Therefore, in order to estimate $\sigma$, $v_{\rm rot}$ and $PA_{\rm kin}$, we adopt a model for the GC kinematics. This approach comes at the cost of assuming that the line-of-sight velocity distribution is Gaussian, but we have shown that this is a reasonable assumption to a first approximation. 

We use the rotation model \citep{Cote01,Proctor}:
\begin{equation}
v_{\rm mod} = v_{\rm sys} \pm \frac{v_{\rm rot}}{\sqrt{1 + \left( \frac{\tan \, (PA - PA_{\rm kin})}{q} \right)^2}}
\label{eq:vmod}
\end{equation}
where $v_{\rm rot}$ is maximum amplitude of the sine curve; $\sigma$ is the standard deviation of the data points with respect to $v_{\rm mod}$; the signs $+$ and $-$ are applied if the PA lies inside or outside the range $(\theta - \theta_0) = [ -\pi/2, +\pi/2]$, respectively; $q=0.84$ is the axial ratio of the GC system, assumed to be the same as that of NGC~4649.

To find the set of $(v_{\rm rot}, \sigma, PA_{\rm kin})$ that best reproduces the data, we minimize the log-likelihood function (i.e. the $\chi^2$ function) \citep{Bergond06,Strader11}:
\begin{equation}
\chi^2 = \sum_i \frac{(v_i - v_{\rm mod})^2}{\sigma^2 + (\Delta v_i)^2} + \ln [\sigma^2 + (\Delta v_i)^2]
\label{eq:chi2}
\end{equation}
where $\sigma$ is the standard deviation of the datapoints with respect to the model.

Uncertainties are derived with ``bootstrapping''. We randomize the sample of interest 1000 times and we fit the data with equations \ref{eq:vmod} and \ref{eq:chi2} each time. We then infer the 68 per cent $(1 \sigma)$ error from the cumulative distribution function of each free parameter. 

\subsection{Independent tracers: stars and planetary nebulae}
\label{sec:tracers}

It is interesting to compare our GC kinematics results with those from two independent kinematic tracers: diffuse starlight from Foster et al. (2015) and planetary nebulae (PNe, \citealt{Teodorescu4649}).

The stellar data extend out to $2 R_e$. We do not compare our results with long-slit or integral-field-unit data because these extend out to $\sim 0.5 R_e$ and they do not overlap with our GC data in the innermost regions. 

The PNe dataset consists of 298 objects. The PN that is furthest from the galaxy centre is at 400 arcsec, which corresponds to the radial extent of the DEIMOS GC sample. We infer the kinematics of PNe using equations \ref{eq:vrms} and \ref{eq:vmod}, as done for GCs. Following \citet{Teodorescu4649}, an average measurement uncertainty $\Delta v = 20 \kms$ is assumed for all PNe. 

The kinematics and spatial distributions of stars and PNe have been shown to be similar \citep{Coccato,Napolitano} because they are different manifestations of the same stellar population. 
It is less clear how the properties of PN systems generally compare with those of the GC system within the same galaxy \citep{Woodley10,Foster11,Forbes2768,Pota13,Coccato13}.

\section{Kinematic results}

We divide our sample into two subsamples. The ``GC'' sample is made by objects with $i>20$ mag. We also require sizes $r_h < 10$ pc for the blue GCs ($(g-z)<1.2$), if available, and $r_h < 20$ pc for the red GCs in order to include faint-fuzzies as part of the red GC sample. This choice is motivated by the findings that faint fuzzies are associated with galaxy disks \citep{ChiesSantos13,Forbes14}, although it is unclear if NGC~4649 possesses an embedded disk, as we discuss in Section \ref{sec:discussion}.  
If physical sizes are not available, the object is still included in the GC sample if the magnitude criterion is satisfied. This GC sample includes classic GCs fainter than $\omega$ Cen, reflecting our attempt to prune from the sample extended objects without size measurements, but it may also include faint extended objects with no size measurements. The ``bright'' sample includes objects with $z<20$ mag with no constraints on the size, and represents bright GCs and UCDs.

Each group is divided into blue and red GCs adopting a conservative colour cut at $(g-z)=1.2$ mag. This choice introduces mixing between the two GC subpopulations for $(g-z)\sim1.2\pm0.1$ mag. Therefore, we also investigate the effects of the colour cut on the results by selecting blue GCs with $(g-z)<1.1$ mag and red GCs with $(g-z)>1.3$ mag, respectively.

We start with a sample of 431 GCs. We exclude one isolated GC at 1500 arcsec from NGC~4649, and one GC because of its proximity to a bright star, which affects the magnitude measurements. In summary, we analyse the kinematics of a sample of 429 objects around NGC~4649.

\label{sec:kineresults}
\subsection{The rms dispersion profile}

We compute the GC $v_{\rm rms}$ velocity dispersion profile in radial bins with roughly 50 objects per bin. For the stars, we compute $v_{\rm rms}^2 = v_{\rm rot}^2 + \sigma^2$, using the tabulated values from Foster et al. (2015).

\begin{figure}
\centering
\includegraphics[width=\columnwidth]{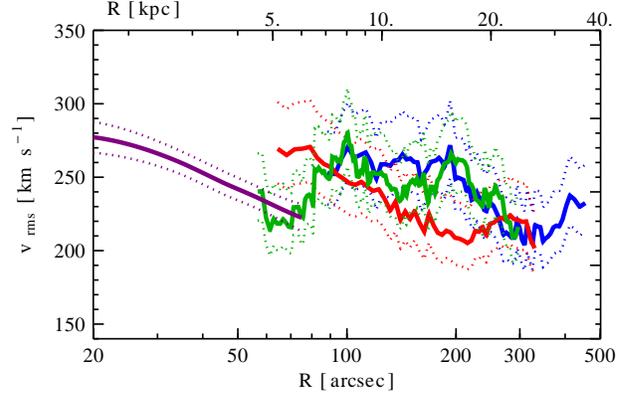} 
\caption{Root-mean-square velocity profile of kinematic tracers in NGC~4649. Shown as think lines are the rolling-binned $v_{\rm rms}$ profiles (equation \ref{eq:vrms}) for blue and red GCs coloured accordingly, respectively. PNe results are shown in green. Stellar data are shown in purple. Dashed lines are $1\sigma$ envelopes. }
\label{fig:vrms}
\end{figure}

We calculate $v_{\rm rms}$ for GCs and PNe in rolling radial bins with 50 objects per bin. The
results are shown in Figure \ref{fig:vrms}. The $v_{\rm rms}$ profiles from the three tracers are diverse. The red GC $v_{\rm rms}$ profile decreases monotonically with radius, whereas the $v_{\rm rms}$ profile of blue GCs and PNe are bumpy and akin to each other, which is unexpected when compared to other galaxies.

We also note that the PNe results agree with the stellar $v_{\rm rms}$ profile in the region of overlap. On the other hand, the results for the red GCs seem to be systematically larger compared to the stars. It is unclear if this effect is real, or is due to unknown systematics in the spectroscopic measurements \citep[e.g.,][]{Arnold}. This result is still significant when a more extreme GC colour cut is adopted.
 
\subsection{Results from the kinematic modelling}

The best fit parameters to eq. \ref{eq:vmod} for GCs and PNe sample (without any radial binning) are given in Table \ref{tab:kin}. GCs from surrounding galaxies are discarded from the analysis. 

\subsubsection{Kinematic radial profiles}
\label{sec:classic}

\begin{figure*}
\centering
\includegraphics[scale=0.85]{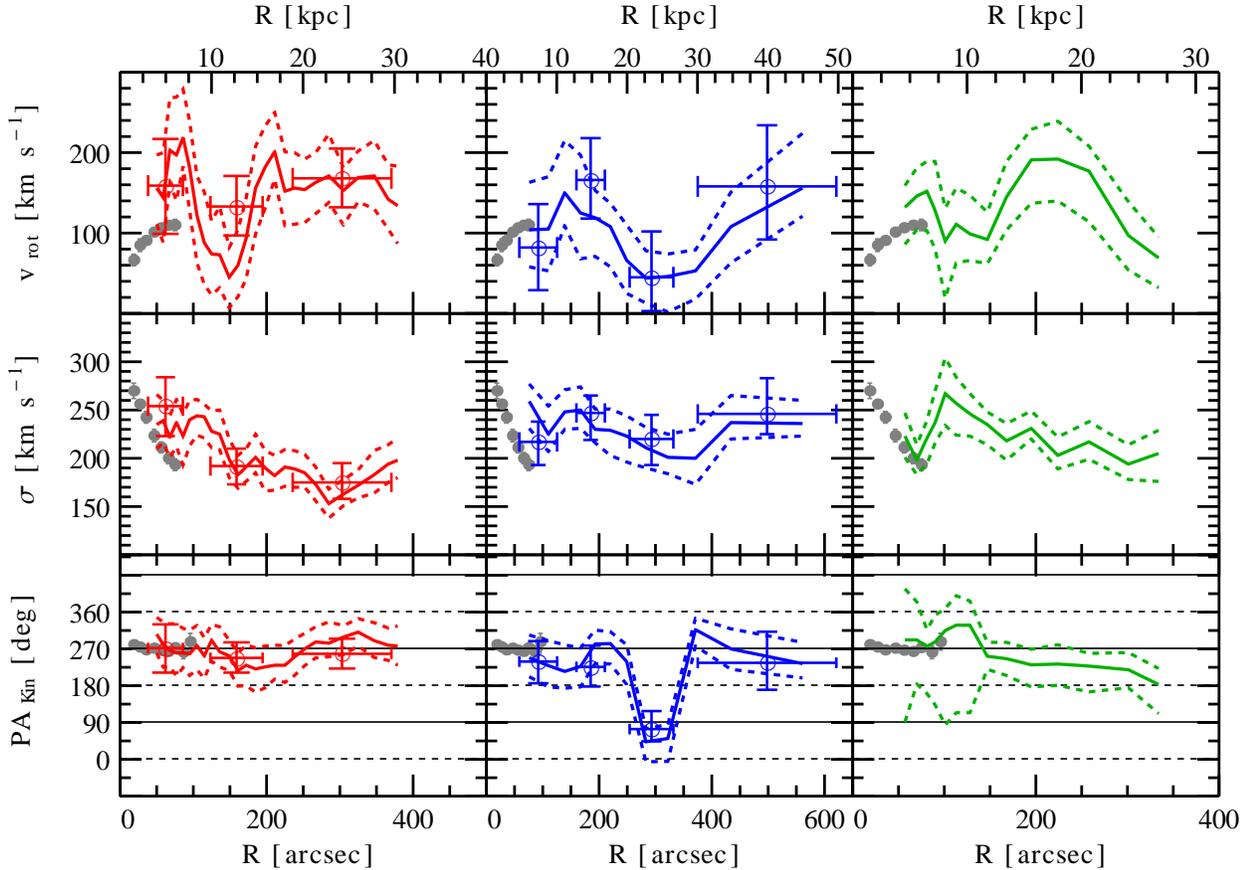} 
\caption{Kinematic radial profiles. Red, blue and green lines represent red GCs, blue GCs and PNe respectively. Open circles are the results for the reddest and the bluest GCs, respectively (see text). Grey points are stellar data from Foster et al. (2015).
Each panel represents a solution to equation \ref{eq:vmod} for a given kinematic tracers. From the top to the bottom we show the rotation velocity $v_{\rm rot}$, the velocity dispersion $\sigma$ with respect to $v_{\rm rot}$, and the kinematic position angle $PA_{\rm kin}$. Rotation occurs near the major axis at $PA\sim270$ deg (marked with a  black line). The dashed line indicates rotation along the minor axis. Running averages are performed with $\sim$ 35, 40, 45 objects bin for red GCs, PNe and blue GCs, respectively. The co-moving step is 5 objects. Rotation along the major axis is detected at all radii. }
\label{fig:kine}
\end{figure*}

Radial kinematic profiles for GCs, PNe and stars, are shown in Figure \ref{fig:kine}. We show running average profiles, with $\sim$ 40 GCs per bin. Focusing on the red GCs (left panels), this sample shows a nearly constant rotation with radius. The rotation axis is very consistent with the photometric major axis of the galaxy ($PA_{\rm kin}\approx270$ deg). The velocity dispersion is bumpy and decreases with radius. The dip in rotation velocity between 100 and 200 arcsec is intriguing and it will require future in-depth analysis. One possibility is that this feature is spurious because it occurs in the radial range where the spiral galaxy NGC~4647 prevents us from detecting GCs, therefore biasing the kinematic fit (see also Figure \ref{fig:pos}). Alternatively, the velocity dip may be real and linked to the recent interaction between NGC~4649 and the galaxy remnant M60-UCD1 \citep{Seth14}, which lies at about 90 arcsec (7.2 kpc) along the minor axis of NGC~4649. 

The blue GCs (middle panels) also show hints of rotation at all radii. The rotation amplitude is lower with respect to the red GCs. As with the red GCs, the rotation direction is generally consistent with the major axis of the galaxy. Minor axis counter-rotation is also detected at 300 arcsec. The velocity dispersion is constant at $\sigma=240\pm30\kms$, and higher than that of the red GCs, as found in other galaxies. 
The GC kinematic results are generally independent of the colour cut, as shown by the open points in Figure \ref{fig:kine}.

\begin{table}
\centering
\label{mathmode}
\begin{tabular}{@{}l c c c c}
\hline
Sample & $N$ & $v_{\rm rot}$ & $\sigma$ &$PA_{\rm kin}$  \\
 &  & [\kms] & [\kms] &[degree]\\ 
\hline
& \multicolumn{4}{c}{GC sample with $(g-i)<1.2$ and $(g-i)>1.2$} \\
\hline
All & 366 & $95_{-17} ^{+18}$ & $229_{-8} ^{+7}$ & $256_{-11} ^{+10}$\\ 
Blue & 212 & $81_{-30} ^{+20}$ & $237_{-10} ^{+9}$ & $244_{-12} ^{+17}$\\ 
Red & 154 & $113_{-20} ^{+23}$ & $216_{-10} ^{+8}$ & $269_{-13} ^{+15}$\\ 
\hline
& \multicolumn{4}{c}{GC sample with $(g-i)<1.1$ and $(g-i)>1.3$} \\
\hline
All & 279 & $115_{-24} ^{+17}$ & $229_{-8} ^{+8}$ & $246_{-9} ^{+13}$\\ 
Blue & 170 & $88_{-28} ^{+28}$ & $238_{-11} ^{+12}$ & $231_{-11} ^{+14}$\\ 
Red & 109 & $150_{-26} ^{+28}$ & $209_{-14} ^{+15}$ & $260_{-11} ^{+14}$\\ 
\hline
& \multicolumn{4}{c}{Bright sample with $(g-i)<1.2$ and $(g-i)>1.2$} \\
\hline
All & 60 & $68_{-37} ^{+57}$ & $273_{-19} ^{+22}$ & $271_{-45} ^{+56}$\\ 
Blue & 35 & $61_{-42} ^{+55}$ & $261_{-43} ^{+35}$ & $144_{-68} ^{+72}$\\ 
Red & 25 & $243_{-66} ^{+55}$ & $238_{-38} ^{+32}$ & $293_{-17} ^{+15}$\\ 
\hline
& \multicolumn{4}{c}{PNe sample} \\
\hline
PNe & 298 & $98_{-20} ^{+17}$ & $234_{-9} ^{+10}$ & $267_{-14} ^{+11}$\\
\hline
\end{tabular}
\caption{Results of the kinematic modelling. Listed are the best fit parameters to eq. \ref{eq:vmod} for different sub-samples, each containing $N$ objects. }
\label{tab:kin} 
\end{table}

Lastly, the PNe (right panels) show  a nearly constant rotation profile, but the kinematic position angle twists with radius, as found by \citet{Coccato13}. The velocity dispersion profile overlaps nicely with the stellar results in the innermost regions. We see a $\sigma$-bump at 100 arcsec, and a declining velocity dispersion profile intermediate between the results for the blue and red GCs.

Figure \ref{fig:kine} shows that the three kinematic tracers are similar to each other in some aspects, e.g., rotation axis and rotation amplitude, but their velocity dispersions behave very differently. 
The sample size of the bright group is too low to compute radial profiles. However, the azimuthally averaged quantities from Table \ref{tab:kin} indicate no significant difference between this group and the results discussed above. 

\subsubsection{Kinematic colour profiles}

\begin{figure}
\centering
\includegraphics[width=\columnwidth]{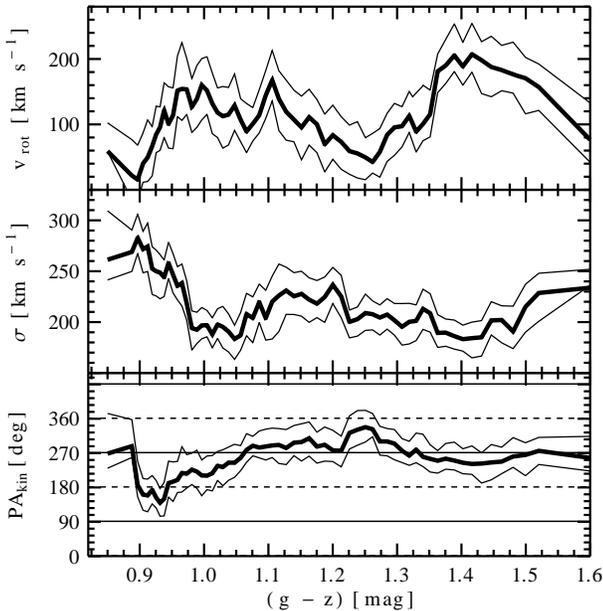} 
\caption{Kinematics as a function of GC colour. Each rolling bin contains 50 GCs. $1\sigma$ envelopes are shown as thin lines. The solid and dotted lines indicate rotation along the major and minor axis, respectively. A colour cut at $(g-z)=1.2$ mag is used to derive the kinematic radial profiles.}
\label{fig:kinecol}
\end{figure}

We have shown that the kinematic modelling results are not very sensitive to the assumed colour cut. We now bin the data by colour, rather than by galactocentric radius, and measure the azimuthally averaged kinematics of the objects within a certain colour bin. This exercise is independent of the colour cut. The results are shown in Figure \ref{fig:kinecol} in the form of co-moving bins with 50 objects per bin. Comparison with the colour GC gradients in Figure \ref{fig:gradients} and Figure \ref{fig:kine} is informative.

The results in Figure \ref{fig:kinecol} show a double-peaked rotation pattern corresponding to the rotation of the blue and red GCs. 
The central panel shows a declining velocity dispersion in the range $0.8<(g-z)<1.0$ mag, which flattens at redder colours. More interestingly, the GCs with $(g-z)>1.55$ mag have a large velocity dispersion and negligible rotation, because they are very centrally concentrated (see Figure \ref{fig:gradients}). The rotation is consistent with the photometric major axis of the galaxy. On the other hand, minor axis rotation is suggested for a group of blue GCs with $(g-z)\sim0.9\pm0.1$ mag, as detected at intermediate radii (Figure \ref{fig:kine}).

\section{Discussion}
\label{sec:discussion}

NGC~4649 is a normal looking elliptical galaxy with a relatively normal GC system.
From a photometric point of view, we have shown that the GC system of NGC~4649 presents typical features seen in other elliptical galaxies of similar mass. We confirm statistically significant GC colour bimodality, as found in earlier work on NGC~4649 \citep{Bridges06,Pierce4649,Lee08,Faifer11,Chies-Santos}. The GC specific frequency within 50 kpc $S_N = 3.5 \pm 0.1$ \citep{Faifer11} is consistent with values obtained for galaxies of similar mass \citep{Peng08}. The two GC subpopulations have different spatial distributions (with the blue GCs being more extended than the red GCs) and negative colour gradients in the innermost regions. 

From a kinematic point of view, the GC system of NGC~4649 is less normal when compared to other systems, because of the significant rotation detected at all radii for both GC subpopulations, which is a rare occurrence in elliptical galaxies of this mass. In what follows, we discuss possible formation scenarios for NGC~4649 based on the results from this paper and additional data from the literature.

\subsection{Comparison with simulations}

\begin{figure}
\centering
\includegraphics[width=\columnwidth]{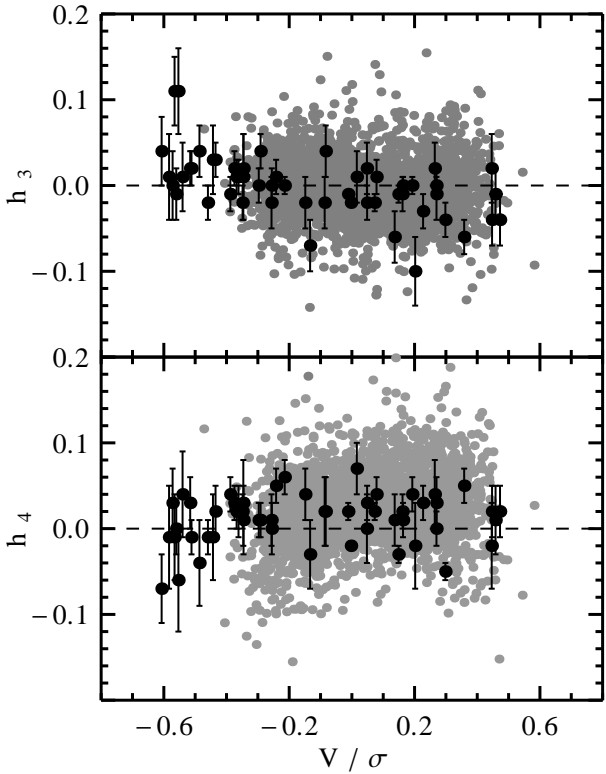} 
\caption{Pixel-by-pixel correlations between $V / \sigma$ and $h_3$ and $h_4$. Gray points are ATLAS$^{\rm 3D}$ data, whereas black points are DEIMOS stellar data within 1 $R_e$.}
\label{fig:h3h4}
\end{figure}

We start by looking at the innermost stellar kinematics in NGC~4649. We use the tools given by \citet{Naab14} to link the observed innermost stellar kinematics from ATLAS$^{\rm 3D}$ with simulated galaxies with known formation histories at $z=0$. 
NGC~4649 is classified as fast-rotator by \citet{Emsellem11}, with parameter\footnote{$\lambda_e$ is a proxy for the galaxy angular momentum within 1 $R_e$ and it is equivalent to the luminosity weighted ratio $(v_{\rm rot}/\sigma)$ \citep{Emsellem07}.} $\lambda_e=0.127$ within 1$R_e$ and ellipticity $\epsilon_e=0.156$. New extended stellar kinematics data from Foster et al. (2015), have shown that the value of $\lambda$ rises up to 0.6 at 2.5 $R_e$, meaning that the galaxy becomes more and more rotation dominated in the outer regions. Moreover, we find that:
\begin{itemize}
\item The pixel-by-pixel values of $V/ \sigma$ (where $V$ is the observed radial velocity) is weakly anti-correlated and correlated $h_3$ and $h_4$, respectively. The Pearson coefficients for the ATLAS$^{\rm 3D}$ data are $\rho_{V / \sigma , h_3} = -0.08$ and $\rho_{V / \sigma , h_4} = 0.24$. This is shown in Figure \ref{fig:h3h4}, where we show both ATLAS$^{\rm 3D}$ data and DEIMOS stellar data from \citet{Foster15} within 1$R_e$. The parameters $h_3$ and $h_4$ describe the shape of the line-of-sight velocity distribution within a given spaxel. The correlation with $V/ \sigma$ is indicative of axisymmetry or of the presence of a disk. 
\item The stellar velocity dispersion is peaked at the centre, and decreases with radius (see Figure \ref{fig:kine}).
\item Isophotes are elliptical (not disky) and there is no evidence of an embedded stellar disc \citep{Kormendy09,Vika12}. 
\end{itemize}

According to the recipe of \citet{Naab14}, galaxies with the above properties formed via one late ($z<1$) gas-poor major merger and many more minor mergers. Major mergers here are meant as those with up to 1:4 mass-to-mass ratio. Less than 18 per cent of the stellar mass formed in-situ, owing to the relatively low amount of gas involved in the merger. The remaining mass fraction, the one which makes up most of the stellar halo, was accreted via minor mergers. The strength of the correlation between $h_3$, $h_4$ and $V / \sigma$ can be linked to the amount of gas involved in the merger \citep{Hoffman09}. However, the observed correlations in Figure \ref{fig:h3h4} are not strong, and therefore we cannot constrain the amount of gas, if any, involved in the major merger.

In principle, the observational properties listed above are also compatible with a gas-rich major merger remnant, but these remnants are usually less massive than $\log M_* = 11.3 M_{\sun}$ and they have disky isophotes, whereas NGC~4649 has $\log M_* = 11.7 M_{\sun}$ \citep{Cappellari13} and no strong evidence for a stellar disk \citep{Vika12}. Spectral energy distribution fitting of NGC~4649 implies an average stellar age of 9.5 Gyr, no recent star formation and negligible gas content \citep{Amblard14}. 

Further evidence supporting a dry major merger formation pathway for NGC~4649 comes from the deficit of stellar mass ($M_{\rm def}$) at the centre \citep{Lauer05,Kormendy09}. The interaction between two supermassive black holes (SMBHs) during a dry merger can evacuate stars from the central regions of the galaxy \citep{Merritt06}. If the mass of the SMBH of the merger remnant is known $M_{\rm SMBH}$, the value of $M_{\rm def}$ can be connected to the number $N$ of dry mergers that a galaxy has undergone, such that $M_{\rm def} / M_{\rm SMBH} \approx 0.5$ per merger. \citet{Dullo14} inferred $M_{\rm def}/M_{\rm SMBH} = 1.22$ for NGC~4649, meaning that this galaxy underwent the equivalent of $\approx$ 2.5 dry major mergers during its assembly, in general agreement with the predicted merger rate of galaxies with comparable stellar mass \citep{Conselice09,Khochfar11,Naab14}.

\begin{figure*}
\centering
\includegraphics[scale=0.8]{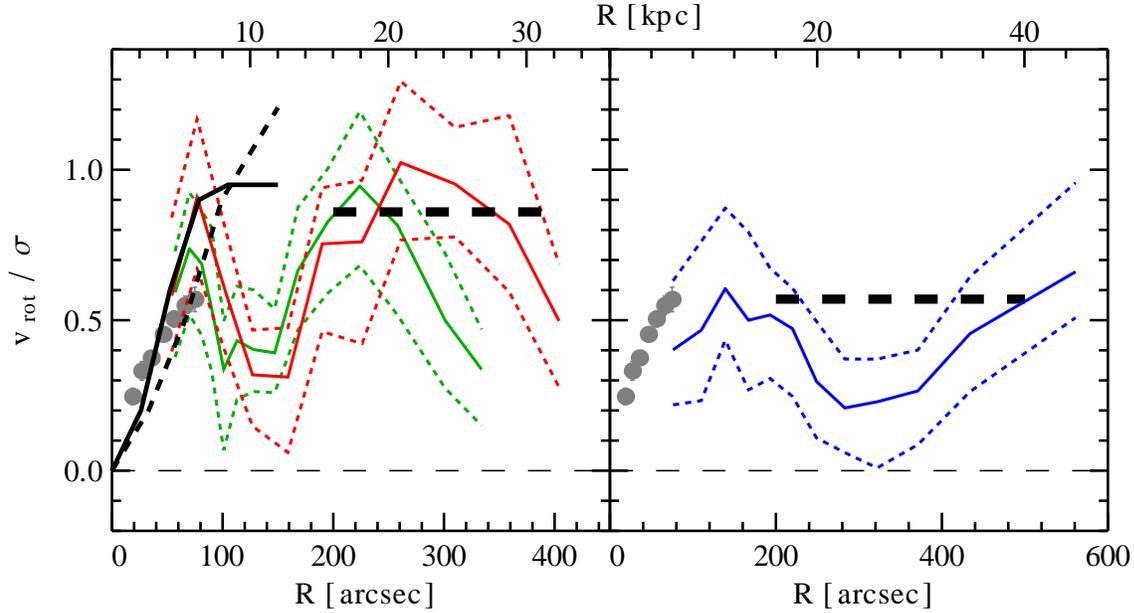} 
\caption{Ratio of the rotation amplitude over velocity dispersion as a function of galactocentric distance. GC data are for the classic GC sample only (i.e. from \S \ref{sec:classic} and Figure \ref{fig:kine}). The left and right panels show the red GCs plus PNe (in green), and blue GCs respectively. Stellar data are shown in both panels as small grey points. The thin dashed line and the thin solid line are the $(v_{\rm rot} / \sigma)$ predicted in a 1:1 dissipationless merger \citep{DiMatteo09} and 1:3 dissipative merger \citep{Cretton}, respectively. Thick dashed lines are the maximum $(v_{\rm rot} / \sigma)$ predicted in the 1:1 dissipationless merger simulations of \citet{Foster11} (model 4 in their Table C1).}
\label{fig:vsigma}
\end{figure*}

In summary, formation via dry major mergers and accretion can qualitatively reconcile the central stellar kinematic profiles with predictions from galaxy formation simulations. We now consider whether this picture is also consistent with the GC constraints discussed in this paper.

As discussed above, rotation in elliptical galaxy outskirts can be interpreted as the dynamical imprint of a major merger between two rotating disk progenitors \citep{Cote01,Cretton,Vitvitska05,Bekki2005,Bournaud07} or between two pressure-supported spheroids \citep{DiMatteo09}. 
The rotation of GCs (and other trace particles) in galaxy outskirts is also expected to have increased after a major merger \citep{Bekki2005,McNeilMoylan12}. This effect can be quantified through the ratio $(v_{\rm rot}/ \sigma)$, which we show in Figure \ref{fig:vsigma} for stars, GCs and PNe. 

The left panel of Figure \ref{fig:vsigma} shows that stars have an increasing $(v_{\rm rot}/ \sigma)$ profile, whereas red GCs and PNe follow a bumpy pattern.
We note that $(v_{\rm rot}/ \sigma)$ is always $<1$, and that the profiles seem to decline outside 20 kpc, which may indicate the turning point of the galaxy rotation curve. On the other hand, the blue GCs (right panel) have less rotation compared to other tracers, with a rising $(v_{\rm rot}/ \sigma)$ profile at large radii.
 
Results for the stars are compared with the predicted $(v_{\rm rot}/ \sigma)$ profile of a gas-rich major merger \citep{Cretton} and of a gas-poor major-merger \citep{DiMatteo09}. Both simulations can reproduce the increasing stellar $v_{\rm rot}/ \sigma$ profile in the region of overlap. The predicted $v_{\rm rot}/ \sigma$ keeps rising with radius, whereas our data show a more irregular rise towards higher $v_{\rm rot}/ \sigma$. 

Similarly, the $v_{\rm rot}/ \sigma$ profile of GCs is compared with the dissipationless major mergers simulations from \citet{Foster11} (thick lines in Figure \ref{fig:vsigma}). These simulations can qualitatively reproduce the maximum $v_{\rm rot}/ \sigma$ of blue and red GCs results for a wide range of orbital configurations (see Table C1 in \citealt{Foster11}). 
Therefore, the radial increase of $v_{\rm rot}/ \sigma$ for the GC is qualitatively consistent with a dry major merger picture. We note that kinematic predictions for GC kinematics in gas-rich major merger remnants are currently not available in the literature \citep{Kruijssen12}. 

\citet{Schauer14} showed that major merger simulations predict that the velocity dispersion profile of the kinematic tracers of the remnant should be ``bumpy'' at intermediate radii. This is a permanent feature within the galaxy, and it has been observed in some systems \citep[e.g.,][]{Pota13}. We find that the velocity dispersion of red GCs and PNe in NGC~4649 is bumpy, especially at $\sim 100$ arcsec (see Figure \ref{fig:kine}). By extrapolating the stellar velocity dispersion profile at large radii, we estimated that the bumps have an amplitude of $\approx$ 50 \kms, in general agreement with the predictions of \citet{Schauer14}.
 
It is known that a galaxy's metallicity profile can be continuously reshaped during its life cycle. Dissipative collapse tends to create steep metallicity radial profiles \citep{Pipino10}, whereas major mergers can steepen or flatten metallicity gradients, depending on the initial conditions of the progenitors \citep{Font}. Similar results can also be achieved through dissipationless major mergers \citep{DiMatteoB}. Other processes, such as AGN feedback, known to have occurred in NGC~4649 \citep{Paggi14}, can flatten the global metallicity profile because they quench star formation in the innermost regions \citep{Hirschmann14}. 

Given these continuous transformations of the metallicity profiles, a direct comparison between the observed colour profiles of stars and GCs with galaxy simulations is more uncertain. Dry major mergers (our favoured formation model so far) can leave the metallicity of the remnant unchanged if both progenitors have sufficiently steep metallicity gradients \citep{DiMatteoB}. In this context, the observed negative gradients in NGC~4649 could be interpreted as the pre-merger gradients of the progenitor galaxies, which remained unchanged after the merger. This assumes that the merger triggered no star formation and that the initial metallicity slopes of the progenitors were equally steep.

Our results show evidence of flat colour profiles outside 20 kpc, consistent with other studies \citep{Rhode,Cantiello07,Bassino,Forbes11,Faifer11,Liu11}. This feature is interpreted as the transition from the galaxy central part, created by dissipative processes, to an extended component made by secular accretion of (mostly metal-poor) satellite galaxies. In this context, GC bimodality is a natural manifestation of dual galaxy formation pathway \citep{Cote99,Brodie14}: red GCs mostly form in-situ, along with bulk of stars, and are therefore more centrally concentrated. Blue GCs are mostly accreted and are therefore more extended \citep{Tonini13}. 

The GC system of NGC~4649 is consistent with being old ($>$ 10 Gyr; \citealt{Chies-SantosAGE}). However, \citet{Pierce06} found about four GC with very young ages in NGC~4649. These objects are listed in Table \ref{tab:young}. From this Table, we can see that young GCs have all red colours, but the sizes are ordinary for normal GCs. Given the lack of recent star formation in NGC~4649, the existence of these young GCs could be explained if they formed along with their host satellite galaxy at $z \le 1$ \citep{Tonini13}, which was then accreted onto NGC~4649.

Lastly, we note that NGC~4649 was classified as a galaxy with ``subtle S0 attributes'' \citep{Sandage81,Sandage94}. Processes such as secular evolution \citep{Larson80,Boselli09,Kormendy13} or gas stripping \citep[e.g.,][]{Moore96,Lake98,Chung07} can both transform a gas-rich spiral galaxy into a gas-poor S0 \citep{Kormendy12}. 

However, galaxies thought to be faded spirals have the photometric stellar disk still in place \citep{Cortesi,Forbes2768,Arnold,Cortesi13,Kormendy13}. No strong evidence for a stellar disk has been found in NGC~4649 \citep{Vika12,Kormendy13}. Recently, \citet{Graham15} pointed out that NGC~4649 is particularly difficult to decompose into bulge and disk.
It was recently shown that gas-rich major mergers can also produce S0 remnants with embedded disks \citep{Querejeta14}. \citet{Naab14} predicts that these remnants have disky kinematics with rising or peaked $v_{\rm rot}/ \sigma$ profiles, along with $h_3$ and $h_4$ strongly correlated with $V/\sigma$, which is not consistent with our findings. 

\begin{figure*}
\centering
\includegraphics[trim = 0mm 20mm 0mm 0mm,clip, scale=0.55]{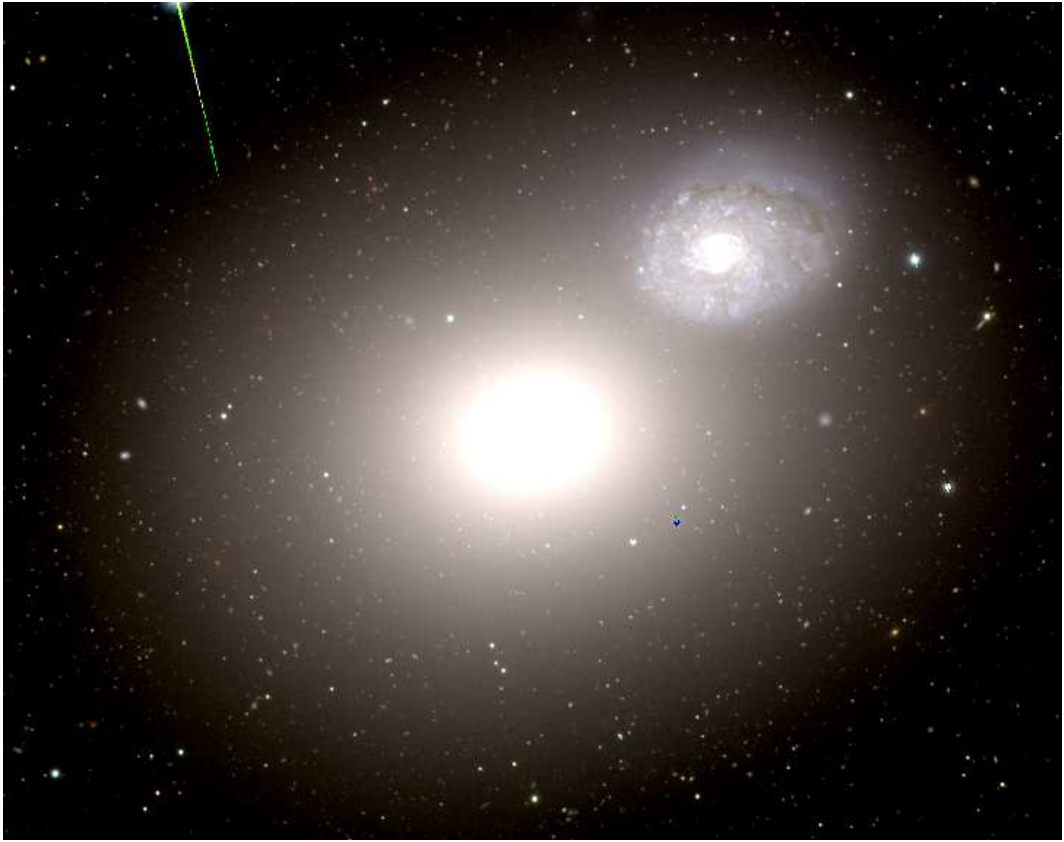} 
\caption{Subaru/Suprime-Cam $gri$ colour image of NGC~4649 (left) and NGC~4647 (right). The projected separation between the two galaxies is 13 kpc. There are no clear signs of interaction between the two galaxies. North is up, East is left. }
\label{fig:colourimage}
\end{figure*}

\begin{table}
\label{mathmode}
\begin{tabular}{l l l l l l}
\hline
ID - ID(P$+$06) & Age & $v$ & $(g-z)$ & $r_h$ & $R$   \\
 & [Gyr] & [\kms] & [mag] & [pc] & [kpc]\\ 
\hline
J187 - 1443 & $2.1\pm0.6$ & $703\pm46$ & $1.53$ & $2.52$ & 3\\  
A8 - 502 & $3.0\pm0.8$ & $705\pm41$ & $1.50$ & $1.87$ & 7 \\  
B36 - 175 & $2.4\pm0.4$ & $836\pm11$ & $1.392$ & $1.95$ & 16 \\  
B85 - 89 & $2.4\pm0.8$ & $1148\pm16$ & $1.29$ & $2.10$ & 18 \\  
\hline
\end{tabular}
\caption{Properties of young GCs in NGC~4649. ID(P$+$06) and age are from \citet{Pierce06}. The ID, colour $(g-z)$ and size $r_h$ are from \citep{Strader12}. The velocity $v$ is from this paper. Errors on $(g-z)$ and $r_h$ are less than 10 per cent. $R$ is the projected distance from NGC~4649. }
\label{tab:young} 
\end{table}

\subsection{Final remarks and the interaction with NGC~4647}

We have shown that a formation pathway via a major merger between two gas-poor galaxies, combined with satellite accretion, can simultaneously explain both the inner stellar observations and the outer GC observations. After a major-merger, the angular momentum is transported to the outer regions of the remnant, explaining the high rotation velocity observed for stars, GCs and PNe.

If no gas was involved in the merger, the GC system of the primordial galaxies got mixed together, preserving their original metallicity gradients. Additional GCs (mostly metal-poor) are then accreted into the halo of the galaxy, explaining the extended spatial distribution of blue GCs. This scenario
is supported by the presence of a depleted stellar core in the galaxy's innermost regions. 

The possible interaction between NGC~4649 and NGC~4647, which is only 13 kpc away in projection, has been widely debated in the literature. A $gri$ color image of this galaxy group is shown in Figure \ref{fig:colourimage}. Radio observations suggest that the morphology of CO and HI maps are mildly disturbed, and so are the corresponding velocity fields \citep{Rubin99,Young06}. However, radio observations alone do not return conclusive results on whether or not this galaxy pair is actually interacting. A similar conclusion was reached by \citet{deGrijs} from an analysis of ACS imaging of this galaxy. Moreover, the distribution of optical light around NGC~4647 and NGC~4649 shows no sign of disturbances (Figure \ref{fig:colourimage}). On the other hand, new observations found evidence for enhanced star formation in the region between NGC~4647 and NGC~4649 \citep{Lanz13,Mineo14}. 

\citet{DAbrusco14} and \citet{Mineo14} have suggested that the substructures in the 2D distribution of GCs from ACS imaging may reflect a recent interaction in NGC~4649, but the role of NGC~4647 is still unclear. No significant sign of ongoing interaction between NGC~4649 and NGC~4647 was observed in the 2D velocity field of the GCs \citep{Coccato13} based on the GC catalogue of \citet{Lee4649}.

Our kinematic results show some interesting features, such as wiggly velocity profiles, counter rotation, and a multi-spin PN system. We find that GCs with $(g-z)\approx0.9\pm0.1$ mag, counter-rotate with respect to main rotation direction of the galaxy, which may represent the imprint of a recent interaction \citep{Seth14}. Moreover, \citet{Pierce06} found a handful of GCs only $2$ Gyr old in NGC~4649, which may be linked to a recent accretion event. Similar kinematic anomalies, such as strong rotation amplitude and multi-spin features, have been found in GC systems of interacting early-type galaxies, such as NGC~4365 \citep{Blom}, and merger remnants like NGC~5128 \citep{Woodley10,Woodley11}. 
Computer simulations may determine whether the kinematic features observed in NGC~4649 encode signatures of an ongoing interaction between NGC~4649 and NGC~4647. 
Based on these contrasting results, it appears that, if NGC~4647 and NGC~4647 are interacting, we are witnessing the very early stages of this interaction. 

%%%%%%%%%%%%%%%%%%%%%%%%%%%%%%%%%%%%%%
%%%%%%%%%%%%%%%%%%%%%%%%%%%%%%%%%%%%%%
%%%%%%%%%%%%%%%%%%%%%%%%%%%%%%%%%%%%%%

\section{Summary}
\label{summary}

In this paper, we have studied the globular cluster (GC) system of the massive elliptical galaxy NGC~4649 (M60). We combined wide-field spectroscopy with wide-field imaging from space-based and ground-based telescopes. 
We build up a photometric catalogue combining space-based \textit{HST}/ACS data from \citet{Strader12}, ground-based archival CFHT/MegaCam, and new Subaru/Suprime-Cam data. GCs are selected based on colour and magnitude constrains. We detect all the photometric features usually seen in GC systems of large ellipticals:
the GC system of NGC~4649 is made up of two subpopulations: the red (metal rich) GCs are centrally concentrated. Their spatial distribution is consistent with that of the starlight and planetary nebulae in NGC~4649. The blue (metal poor) GCs are more extended. The colour gradients of blue and red GCs are flat outside 20 kpc, interpreted as evidence of accretion of metal-poor satellites into the stellar halo of NGC~4649. 

The spectroscopic follow-up is performed with three multi-object spectrographs: Keck/DEIMOS, Gemini/GMOS and MMT/Hectospec. These cover different effective areas around NCC~4649, but they are complementary with each other. 
We confirmed 431 GCs associated with NGC~4649, plus some GCs from surrounding galaxies, and dwarf galaxies. 
The contamination from stars and background galaxies is consistent with zero. We analyse the kinematics of blue (60 per cent of the total sample) and red GCs (40 per cent), along with literature planetary nebulae data. Results are:
\begin{itemize}
\item Significant rotation at all radii is detected for stars, PNe and blue and red GCs. Rotation occurs preferentially along the major axis of the galaxy.
\item The velocity dispersion of the three tracers look very different. Blue GCs and PNe have higher velocity dispersion in the outer regions, whereas the red GCs and the innermost PNe are consistent with the stellar kinematics and have lower dispersions. 
\item We report peculiar kinematic features for both GCs and PNe. The rotation axis of PNe twists with radius, whereas a group of blue GCs at intermediate radii counter-rotate with respect to the bulk of blue GCs. The velocity dispersion profiles are bumpy, a feature which may indicate that this galaxy underwent major mergers in the past.
\end{itemize}

We discuss possible formation scenarios for this galaxy, contrasting stellar and GC observations with recent galaxy formation simulations. We find that a formation via dry (no gas) major-merger between two galaxies can, regardless of their initial angular momentum, consistently reproduce stellar observations in the innermost regions and GC observations in the outermost regions of NGC~4649. 
We consider the possibility that NGC~4649 was spiral galaxy which evolved in an elliptical galaxy because of gas stripping, but the apparent non-detection of the primordial disk in NGC~4649 makes this scenario unlikely. We find no strong evidence to support an upcoming interacting between NGC~4649 and the spiral NGC~4647. 

%%%%%%%%%%%%%%%%%%%%%%%%%%%%%%%%%%%%%%
%%%%%%%%%%%%%%%%%%%%%%%%%%%%%%%%%%%%%%
%%%%%%%%%%%%%%%%%%%%%%%%%%%%%%%%%%%%%%

\section*{ACKNOWLEDGEMENTS}

We would like to thank the referee for the constructive feedback.
DAH acknowledges the support of research funding in the form of a Discovery Grant through the Natural Sciences and Engineering Research Council of Canada (NSERC). This research has made use of the NASA/IPAC Extragalactic Database (NED) which is operated by the Jet Propulsion Laboratory, California Institute of Technology, under contract with the National Aeronautics and Space Administration.  This work was supported by NSF grant AST-1211995. We thanks the ARC for financial support via DP130100388.

This research made use of Montage, funded by the National Aeronautics and Space Administration's Earth Science Technology Office, Computation Technologies Project, under Cooperative Agreement Number NCC5-626 between NASA and the California Institute of Technology. Montage is maintained by the NASA/IPAC Infrared Science Archive.
Some of the data presented herein were obtained at the W. M. Keck Observatory, operated as a scientific partnership among the California Institute of Technology, the University of California and the National Aeronautics and Space Administration, and made possible by the generous financial support of the W. M. Keck Foundation. The authors wish to recognise and acknowledge the very significant cultural role and reverence that the summit of Mauna Kea has always had within the indigenous Hawaiian community. The analysis pipeline used to reduce the DEIMOS data was developed at UC Berkeley with support from NSF grant AST-0071048. Based in part on data collected at Subaru Telescope and obtained from the SMOKA  (which is operated by the Astronomy Data Centre, National Astronomical Observatory of Japan), via a Gemini Observatory time exchange. The authors acknowledge the data analysis facilities provided by IRAF, which is distributed by the National Optical Astronomy Observatories and operated by AURA, Inc., under cooperative agreement with the National Science Foundation. We have used the data products from the 2MASS, which is a joint project of the University of Massachusetts and the Infrared Processing and Analysis Centre/California Institute of Technology, funded by the National Aeronautics and Space Administration and the National Science Foundation.

\bibliographystyle{mn2e}
\bibliography{ref}

\end{document}